\documentclass[fleqn,usenatbib]{mnras}
\usepackage{newtxtext,newtxmath}
\usepackage[T1]{fontenc}

\DeclareRobustCommand{\VAN}[3]{#2}
\let\VANthebibliography\thebibliography
\def\thebibliography{\DeclareRobustCommand{\VAN}[3]{##3}\VANthebibliography}

\usepackage{graphicx}	
\usepackage{amsmath}	
\usepackage{subcaption}
\usepackage{xcolor}

\usepackage{orcidlink}

\title[Dynamic galaxy decomposition]{Probabilistic model for dynamic galaxy decomposition}
 
\author[Y. Jagvaral et al.]{Yesukhei Jagvaral\thanks{E-mail:  yjagvara@andrew.cmu.edu}$^1$\orcidlink{https://orcid.org/0000-0001-7068-7037}, 
Duncan Campbell$^1$, 
Rachel Mandelbaum$^1$\orcidlink{0000-0003-2271-1527}, 
\newauthor 
Markus Michael Rau$^1$\orcidlink{https://orcid.org/0000-0003-3709-1324}
\\
$^{1}$Department of Physics, McWilliams Center for Cosmology, Carnegie Mellon University, Pittsburgh, PA 15213, USA }

\date{Accepted XXX. Received YYY; in original form ZZZ}

\pubyear{2020}

\begin{document}
\label{firstpage}
\pagerange{\pageref{firstpage}--\pageref{lastpage}}
\maketitle

\begin{abstract}
In the era of precision cosmology and ever-improving cosmological simulations, a better understanding of different galaxy components such as bulges and discs will give us new
insight into galactic formation and evolution. 
Based on the fact that the stellar populations of the constituent components of galaxies differ by their dynamical properties, we develop two simple models for galaxy decomposition using the IllustrisTNG cosmological hydrodynamical simulation. 
The first model uses a single dynamical parameter and can distinguish 4 components: thin disc, thick disc, counter-rotating disc and bulge. The second model uses one more dynamical parameter, was defined in a probabilistic  manner, and  distinguishes  two components: bulge and disc. The number fraction of disc-dominated galaxies at a given stellar mass obtained by our models agrees well with observations for masses exceeding $  \log_{10}(M_*/M_\odot)=10$.
S\'ersic  indices and half-mass radii for the  bulge components agree well with those for real galaxies. 
The mode of the distribution of S\'ersic  indices for the disc components is at the expected value of $n=1$. However, disc half-mass radii are smaller than those for real galaxies, in accordance with previous findings that the IllustrisTNG simulation produces undersized discs. 
\end{abstract}



\section{Introduction}

Galaxies exhibit a wide range of morphological features. Galaxy morphology carries information about the galaxy's evolutionary history; morphology strongly correlates with color, star-formation rate, the local environment and numerous intrinsic properties   \citep{book, van_den_cent_gal,determines_size,size_velo_early,mass_local,color_mass_stellar,morph_color_color,lum_size}. For instance, in bulge-dominated galaxies, the motions of stars are dispersion-dominated, while  in disc-dominated galaxies, the stellar motions are angular momentum-dominated. Nevertheless, there are still open questions about galaxy formation and evolution, and the connection to galaxy morphology \citep{theory_challenge, gal_problems}. For example, does the environment play a dominant role in setting the morphology, or does assembly bias pre-determine the morphology \citep{zoo_env}? Furthermore, galaxy color is known to correlate with the magnitude of large-scale intrinsic alignments \citep{hirata2007}, which are an important contributor source of systematic uncertainty for weak lensing \citep{2016MNRAS.456..207K}.  Explanations for this color trend generally rely on the color-morphology correlation, identifying red galaxies as dispersion-dominated and blue galaxies as angular momentum-dominated \citep{heymans,2015SSRv..193...67K}. Large-volume cosmological simulations of galaxy formation provide an opportunity to study these questions \citep{croft2009,bluetides2015, eagle2020}.
 
However, in order to study galaxy morphology in a cosmological context, a method that can efficiently and accurately categorize galaxies and decompose them into their constituent structural components at moderate resolution is needed. 
Historically, observed galaxies were classified by their visual characteristics.  
Galaxy morphologies were first categorized by Edwin Hubble in 1926; in particular he identified 4 main categories: elliptical, disc, lenticular and irregular galaxies. 
Additionally, Hubble identified substructures of disc galaxies and drew parallels between the appearances of bulges of disc galaxies and elliptical galaxies. 
In the following decades, beyond just morphology, observations began to reveal the different characteristics of the light intensity profiles of galaxy components.  De Vaucouleurs  introduced
the $r^{-1/4}$ law for describing the light intensity profile of elliptical galaxies and spheroidal components of disc galaxies  \citep{de1948}. Later in 1959, De Vaucouleurs showed that for M31, the surface brightness profile is a sum of an exponentially decaying disc and a bulge that follows the $r^{-1/4}$ law, setting a precedent for all later photometric studies \citep{de1958}.
 Later studies found additional structures such as pseudo-bulges, bars, stellar halo and thick discs  \citep{thickdisk, pseudobulge, hst_halo}, which pose some challenges for photometric fits. 

More recently, studies using resolved spectroscopic information have enabled galaxies to be decomposed into components based on their dynamics, rather than surface brightness profiles  \citep[e.g.,][]{zhu-califa,tabor-manga}.
These studies rely on the assumption that elliptical galaxies and bulge components in disc galaxies are  similar in their formation and evolution. Both are usually red in color, quiescent, and dynamically dispersion-dominated. On the other hand, discs are usually blue, star forming and dynamically rotation-supported \citep{Somerville2014}.

On the theoretical side, up until very recently cosmological hydrodynamical simulations had difficulty producing disc-dominated galaxies that have properties consistent with those observed in real data \citep{vogelsberger-review}.  
 The improved numerical methods and updated implementations of baryonic physics in the IllustrisTNG100 \citep{ tng-bimodal,pillepich2018illustristng, Springel2017illustristng, Naiman2018illustristng, Marinacci2017illustristng,tng-publicdata} simulations produce optical morphologies of galaxies that compare favorably to observed galaxies, thus allowing a reasonable split into disc and elliptical galaxies \citep{tngimage2019}. Additionally, the fraction of disc galaxies from the IllustrisTNG100  simulation  agrees with observations at $z=0$ and stellar mass range $\log_{10}(M_*/M_\odot) = 9.5 - 11.5$ \citep{tacchella2019}. 
 These properties of the IllustrisTNG100 simulation will enable studies of the ensemble statistics of the different galaxy components, from which we can get a hint into their evolutionary history. For instance, understanding how discs assemble and evolve is key to understanding galaxy evolution as a whole, since discs carry a significant portion of the angular momentum of the galaxy \citep{vanderKruit}.

The different components of disc galaxies are thought to form and evolve by different mechanisms \citep{hierarchical}. 
Therefore, comparing the ensemble statistics of different components may reveal new insight into the formation and evolution of galaxies.
Moreover, elliptical galaxies and disc galaxies are known to exhibit different behaviors; for example, intrinsic alignments of galaxy shapes with the large-scale density field have been detected for red galaxies, which are predominantly elliptical, but not for blue galaxies, which are predominantly disc-dominated  \citep{joachimi_2015}. The different kinematic mechanisms in the evolution of the two types of galaxies are thought to be responsible for the difference \citep{hirata2004}.  
Several methods for  dynamical decomposition of galaxies have been demonstrated in the literature, mostly on the outputs of cosmological zoom-in simulations 
These methods have gradually improved over time, building on lessons learned from previous dynamical morphological classifiers. 
\cite{Abadi_2003} first proposed the widely used \textit{circularity} parameter, which characterizes each star particle's orbit and orientation in the reference plane defined by the net angular momentum of the galaxy. Later,
\cite{Scannapieco_2009} studied eight galaxies from a high-resolution hydrodynamical simulation. This work found the circularity parameter to be inadequate for a detailed study, and proposed an additional cutoff based on distance from the center of the galaxy. 
Next, 
\cite{domenech2012} examined the photometry (color) and chemistry of the components of seven disc galaxies in a zoom-in simulation. The decomposition into substructures was done using k-means clustering of various kinematic parameters.  
Improving upon that method, \cite{obreja2018} used Gaussian Mixture Models on dynamical parameters of galaxies from zoom-in simulation. 
Later, \cite{Du_2019} used the same method on unbarred disk-dominated galaxies of IllustrisTNG100, demonstrating that multiple galactic substructures can be found in large-volume simulations as well.  
Additionally, \cite{brook2012} studied the chemical composition and evolution of the bulge fraction over time in a zoom-in  simulation. Instead of kinematics, they used chemical composition to distinguish the thin disc, thick disc and halo. 

However, these morphological decomposition methods cannot be applied to ensembles of galaxies with a variety of morphologies, since they were specifically designed for disc-dominated galaxies. Hence, when working with large samples of galaxies, studies often include a single parameter with a hard cut-off  to separate the populations \citep{vogelsberger_2014, Tenneti_2016,  bluetides,tacchella2019}. These cut-offs do not capture the underlying complexities of the substructures; as a consequence, the separated populations suffer from considerable contamination \citep{Scannapieco_2009}.  
In order to compensate for this problem,
\cite{tacchella2019} used for each galaxy the fraction of the angular momentum in the azimuthal direction in IllustrisTNG100 
to study galaxy morphologies, and applied a global correction factor for the spheroidal component by arguing that this parameter underestimates the spheroidal component.  

Since the aforementioned decomposition models required pre-selected sample of disc-dominated galaxies, our goal is to develop a more general dynamic model for morphological composition of mixed galaxy ensembles. This model should have a relatively small number of parameters, to enable its application to  ensembles of galaxies from cosmological hydrodynamical simulations (not just zoom-in simulations). We test the model presented in this work by comparing it to current observational data and previous morphological decomposition methods.

Also, observational morphological decomposition models conventionally do not decompose the halo stars into a separate structure from the bulge. Even though the halo stars are a separate population from the bulge, they usually constitute a very small fraction of the mass content of the galaxy. Thus, in order to be consistent with observational models, we do not attempt to identify a structure for the halo stars. 

The paper is organized as follows: In \S~\ref{simulation} we briefly introduce the simulation we have used: IllustrisTNG.
Next, in \S~\ref{methods}, we discuss the dynamical parameters that are relevant for morphological decomposition of galaxies, and introduce our 1D and 2D models for morphological decomposition. In \S~\ref{results} we present and discuss ensemble statistics of the galaxy population in IllustrisTNG100 derived using the model. Finally, we summarize our conclusions in \S~\ref{conc}.
 
\section{Simulated data}\label{simulation}

In this section, we briefly describe the IllustrisTNG simulation used throughout this work \citep[for more information, please refer to][]{ tng-bimodal,pillepich2018illustristng, Springel2017illustristng, Naiman2018illustristng, Marinacci2017illustristng,tng-publicdata}.
The IllustrisTNG100-1 is a cosmological hydrodynamical simulation that was run with the moving-mesh code Arepo \citep{arepo}. The box (with side length of 75 Mpc/h
) has $2 \times 1820^3$ resolution elements with a gravitational softening length of 0.7 kpc/h 
for dark matter and star particles.  For gas cells, an adaptive comoving softening of a minimum value of $\sim 0.2$ kpc is used.
The dark matter and star particle masses are $7.46\times 10^6  M_\odot$ and $1.39\times 10^6  M_\odot$, respectively.  The physical models and processes for galaxy formation and evolution include radiative gas cooling and heating; star formation in the ISM; stellar evolution with metal enrichment from supernovae; stellar, AGN and blackhole feedback; formation and accretion of supermassive blackholes \citep{tng-methods}.
The halos within the simulation are identified through friends-of-friends (FoF) 
methods \citep{fof}, and then the subhalos  
are identified using the SUBFIND algorithm \citep{subfind}. 

We chose to use this simulation for this work for the following reasons. The $g-r$ galaxy colors in the IllustrisTNG100-1 simulation (the highest resolution publicly available data set)  are in quantitative agreement with observational data from SDSS at $z < 0.1$. 
The simulation exhibits improved
color bimodality that agrees  with SDSS data for intermediate mass galaxies compared to previous simulations, such as Illustris \citep{tng_2color}. 
Further, the $g-r$ colors for the galaxies are correlated with other galaxy properties; for example, the red galaxies are quiescent, old,  gas poor and less rotationally-supported. Finally, the cosmological nature of the simulation enables us to derive statistical properties of galaxy ensembles, unlike zoom-in simulations.  
 
Of the 100 snapshots of the simulation, we use the latest snapshot at $z=0$ for our analysis.  We initially employ a minimum stellar mass threshold of $ \log_{10}(M_*/M_\odot) =9 $ , roughly corresponding to $10^3$ star particles \citep{ Tenneti_2016,Du_2020}.  However, the results of our analysis in \S~\ref{n_frac}  
motivate a stricter cut of $ \log_{10}(M_*/M_\odot) =10 $  in order to avoid resolution limitations in calculation of derived quantities such as galaxy sizes for individual galaxy components.  In order to compare our measured galaxy properties with observations, we employ a cut of five times the 3D half-mass radius for all measurements, similar to \cite{tacchella2019,Du_2020}. We have confirmed that this radial mask has very little effect (less than 0.1\%) on measured quantities such as the fraction of stars in the disc. 
\begin{figure*}
\begin{center}
$\begin{array}{c@{\hspace{0.5in}}c}
\includegraphics[width=7.2in,angle=0]{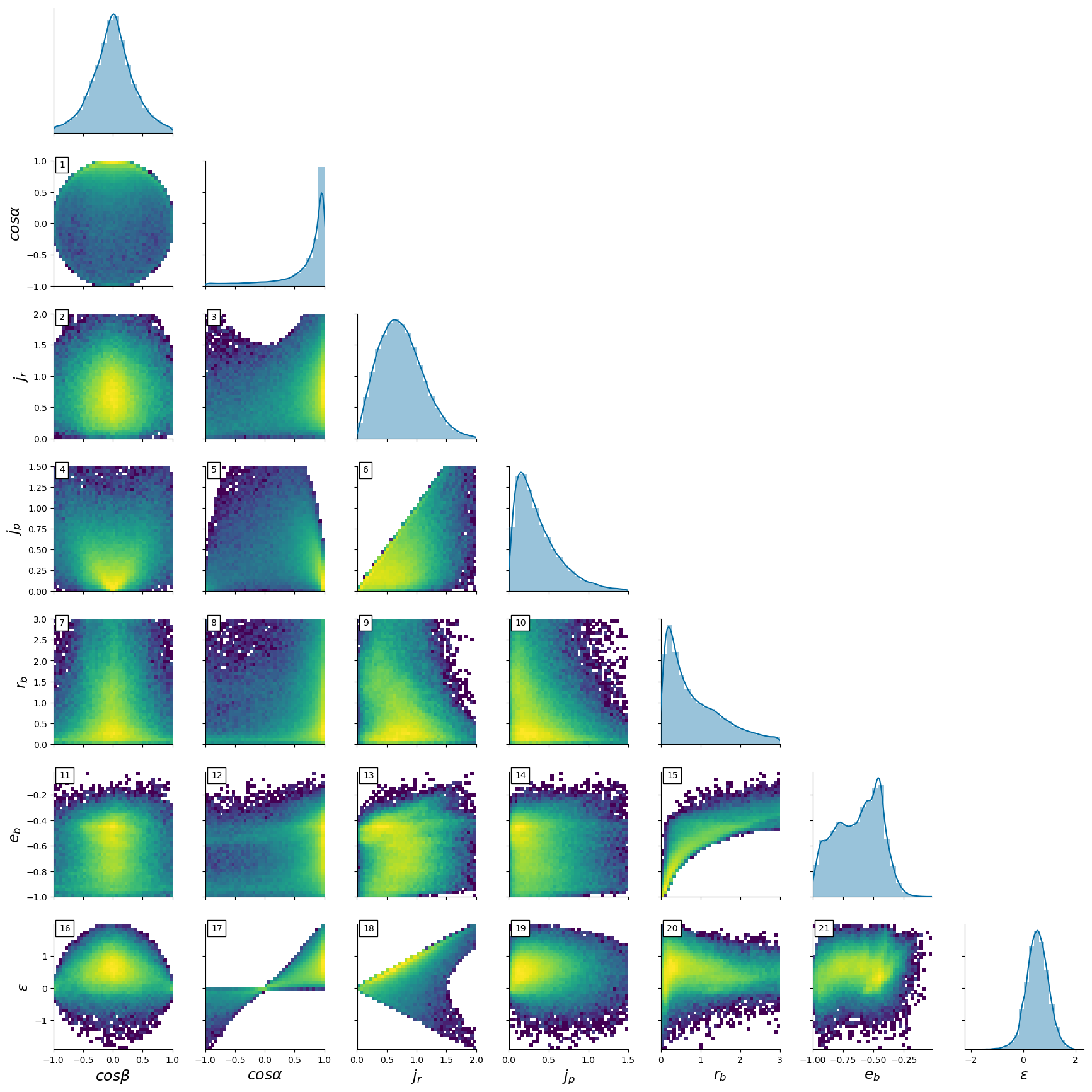} 
\end{array}$
\caption{ \label{corner} Corner plot of the kinematic phase space of 7 parameters described in \S~\ref{kin_params} for one typical disc-dominated galaxy of mass $ \log_{10}(M_*/M_\odot)\sim 10 $. The logarithmic color scale is common across the panels. For description and discussion about the parameters please refer to \S~\ref{kin_params} of the text.  
}
\end{center}
\end{figure*}

\section{Methods}\label{methods}
In this section, we introduce dynamical parameters that were considered for the morphological decomposition model, and discuss their suitability for such a model. We then discuss several potential kinematic decomposition methods that will be used in this work, along with the statistical methods used to discriminate between models.

\subsection{The kinematic phase space of the stars}\label{kin_params}

In order to dynamically decompose a galaxy,  we consider relevant parameters of the phase space. We expect stars belonging to the same structural component to cluster together in the dynamical phase space. The correlation and dependence of these parameters in the phase space will reveal the underlying physical structure of the galaxy. For each galaxy we take the most bound particle of the subhalo as the origin of the spatial coordinate system, and the direction of the galaxy's total angular momentum as $\hat{\mathbf{z}}$. 

First, in an effort to derive physically meaningful information from various  combinations of angular momentum-related parameters that enable classification into different morphological components
, we define the following parameters:
\begin{itemize}

   \item $\mathbf j_\text{star}$ is the angular momentum of a single star particle and $j_\text{star}$ is its magnitude. 
\item  $j_\text{circ}(r) = r\, v_\text{circ}(r) = \sqrt{\frac{GM(r)}{r}}$ is the expected angular momentum for a circular orbit at the same position as that star,
where $M(r)$ is the total mass (across all types of particles -- stars, gas, dark matter) contained within that radius. 
 \item $   j_\text{p} = (j_\text{star} - \mathbf j_\text{star}  \cdot  \hat{\mathbf{z}}) / j_\text{circ}$ is the magnitude of the angular momentum of a single star particle that is not aligned with the galaxy's angular momentum scaled by $j_\text{circ}$. This parameter should be close to zero for disc stars.  
  \item $ \varepsilon = \frac{\mathbf j_\text{star}  \cdot  \hat{\mathbf{z}}} {j_\text{circ}(r)}$ is the circularity parameter that has been used widely in literature, since \cite{Abadi_2003}. Disc stars will tend to be concentrated around $ \varepsilon \sim 1$. We note that the original definition of \cite{Abadi_2003} used $j(E)$ (the maximum specific angular momentum possible
at the specific binding energy $E$ of the star particle), rather than  ${j_\text{circ}(r)}$. However,  these two definitions yield very similar results \citep{marinacci}.
  \item  $j_\text{r} \equiv \frac{j_\text{star}} {j_\text{circ}(r)}$ indicates the type of orbit taken by the star particle. Stars on circular orbits will have $j_\text{r} \sim  1$, while those on elliptical orbits can have $j_\text{r}$ either above or below 1.
  \end{itemize}
  
Galaxy components may exhibit patterns not only in the dynamical phase space, but also in configuration space.  We therefore introduce a second set of parameters quantifying the star particle positions relative to the galactic plane and center:
\begin{itemize}
 \item $r_\text{b} = r/R_\text{1/2}$, where $r$ is the 3D radial distance of the star from the origin and $R_\text{1/2}$ is the half-mass radius.  Star particles associated with the disc may extend further from the origin than the bulge stars.
  \item $\cos\beta$ is the cosine of the angle between the position vector of the star particle and the total angular momentum of the galaxy.  Since disc particles will tend to lie in the plane normal to the galaxy angular momentum vector, a concentration of star particles at $\cos\beta \sim 0$ signals that the galaxy contains a disc structure.
  \item $\cos\alpha$ is the cosine of the angle between the angular momentum vector of the star particle and the total angular momentum of the galaxy. A concentration of particles at $\cos\alpha \sim 1$  signals that the galaxy contains a disc structure, since particles preferentially have their angular momentum aligned with that of the galaxy overall. Spread within the two angular parameters $\cos\alpha$ and $\cos\beta$ signifies the disordered motion among bulge stars. 
    
  \cite{sales_2010} proposed the use of $\kappa_\text{rot} = K_\text{rot}/K$, where $K$ is the total kinetic energy of the star and $K_\text{rot} = m/2 ( \mathbf j_\text{star}  \cdot  \hat{\mathbf{z}} /r)^2$ is the rotational kinetic energy.  However, this quantity is equivalent to  $\cos^2\alpha$, consequently neglecting the counter-rotating particles altogether.  We therefore omit it due to its redundancy once we use $\cos\alpha$.  
    \end{itemize}

Finally, we consider a single energy-related parameter:
\begin{itemize}
  \item $e_b$ is the specific binding energy of the star scaled by absolute binding energy of the most bound stellar particle in the subhalo, $|e|_\text{max}$,    
  as defined in \cite{obreja2018}.  
  $e_b$ should  exhibit a strong correlation with distance, due to trends in the gravitational potential with distance. Consequently, bulge star particles will have lower binding energies compared to disc star particles, making this parameter a viable option for bulge-disc separation.
   
  \end{itemize}

In Fig.~\ref{corner} we present a corner plot of these parameters for a typical disc-dominated  
galaxy of mass $ \log_{10}(M_*/M_\odot) \sim 10 $. 
The 1D histogram of  $\cos\alpha$ exhibits a high concentration of stars approaching $\cos\alpha \sim 1 $, whereas  the angle $\beta$ exhibits a concentration of stars around $\cos\beta \sim 0 $.  
$\alpha$ is a kinematic angle that also contains information about the star particle angular momentum direction with respect to the galaxy angular momentum direction, which is useful for studying counter-rotating discs and disordered motion.
For this reason, we chose $\cos\alpha$ for decomposition.  

Next are the set of parameters relating to angular momenta of the star particles.
Stars with $j_\text{r} \sim 1$ 
are on circular orbits, while stars with elliptical orbits exhibit a wide range of $j_\text{r}$ values, including ones that are both near and far from 1. 
As a result, we  may use this parameter to distinguish particles based on their orbital shape.  
 The $j_\text{p}$ parameter should be close to zero for disc stars and should essentially carry the same information for bulge stars as $j_\text{r}$.  In particular,  $j_\text{p}$ encodes information about motion in the equatorial plane of the galaxy, but this information is already captured by the $\cos\alpha$ parameter.
 Also, $j_\text{p}$ does not exhibit any bimodality corresponding to different components, which suggests that it does not add significant information for morphological decomposition.  
 
The circularity parameter $\varepsilon$ has been used with a constant cutoff of 0.7 to distinguish between the bulge and the disc components of simulated galaxies \citep{marinacci,Tenneti_2016}. As illustrated in Fig.~\ref{mc3},  it arbitrarily and severely undercuts the disc component, whilst also contaminating the disc with bulge stars. In order to overcome this shortcoming, \cite{Scannapieco_2009} proposed to additionally use distance $r$, as shown in Fig.~\ref{corner} subfig.~20. 
However, the methodology of \cite{Scannapieco_2009} requires two-component galaxies to exhibit two distinct populations in the plane of  $\varepsilon$ and $r_b$.  
Neither the specific galaxy shown in Fig.~\ref{corner} nor the broader galaxy population in this simulation exhibits this feature.  For this reason, we do not use the combination of $\varepsilon$ and $r_b$  for morphological decomposition. 
 
In addition, two-population separation (or bimodality in 1D) is also not common among  two-component galaxies in $e_b$. In addition, Fig.~\ref{corner} subfig.~15 shows that distance and binding energy are highly correlated as expected, consequently the aforementioned arguments for $r_b$ applies to $e_b$ as well.  
Another reason that binding energy may not be an optimal choice of a parameter for identifying dynamical substructures is that smaller substructures such as spiral arms, bars and globular clusters may complicate the distribution, since these are regions of high stellar concentration.  

In summary, after considering a wide range of parameters quantifying the spatial and kinematic distributions of star particles within galaxies, we investigate kinematic decomposition based on $\cos\alpha$, and later add $j_\text{r}$, as it provides additional information about the orbital shape. 

\subsection{1D kinematic decomposition model}
 
We begin our exploration of kinematic decompositions as simply as possible, with methods based on a single dynamical parameter from amongst those discussed in \S~\ref{kin_params}.  Before developing new methodology, we illustrate how a commonly-used 1D decomposition method works in IllustrisTNG.
 
\begin{figure*} 
\begin{subfigure}{.3\textwidth}
\centering
\includegraphics[width=5.5cm]{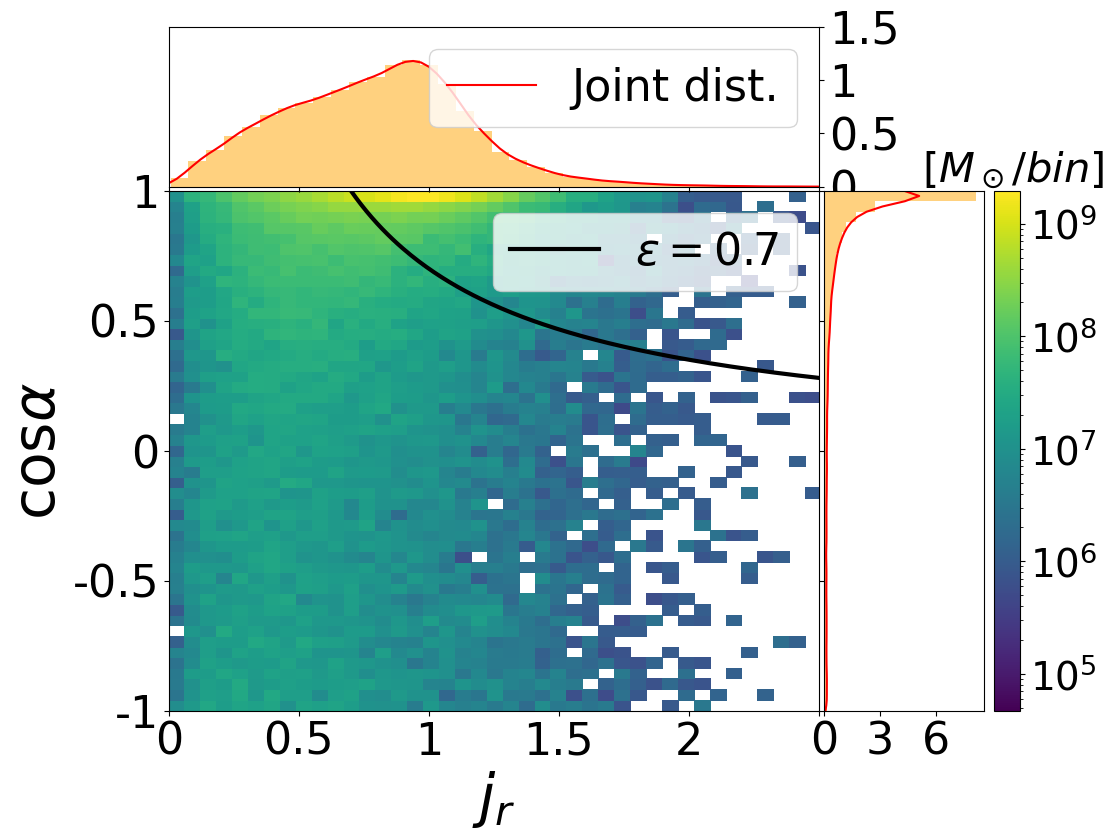}
\caption{\label{mc3_a}The whole galaxy before Monte Carlo}
\end{subfigure}\hfill
\begin{subfigure}{.3\textwidth}
\centering
\includegraphics[width=5.5cm]{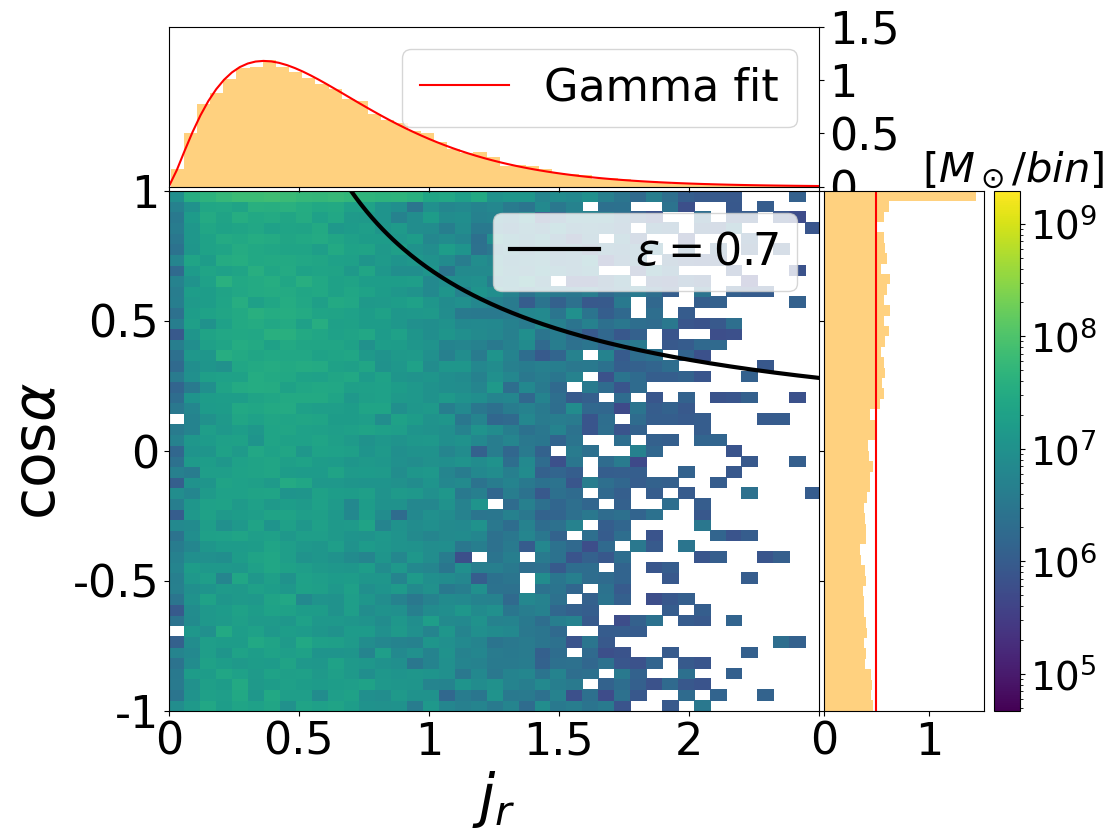}
\caption{\label{mc3_b}The bulge component after Monte Carlo; note uniformity in $\cos\alpha$, compare with Fig.~\ref{ellip} }
\end{subfigure}\hfill
\begin{subfigure}{.3\textwidth}
\centering
\includegraphics[width=5.5cm]{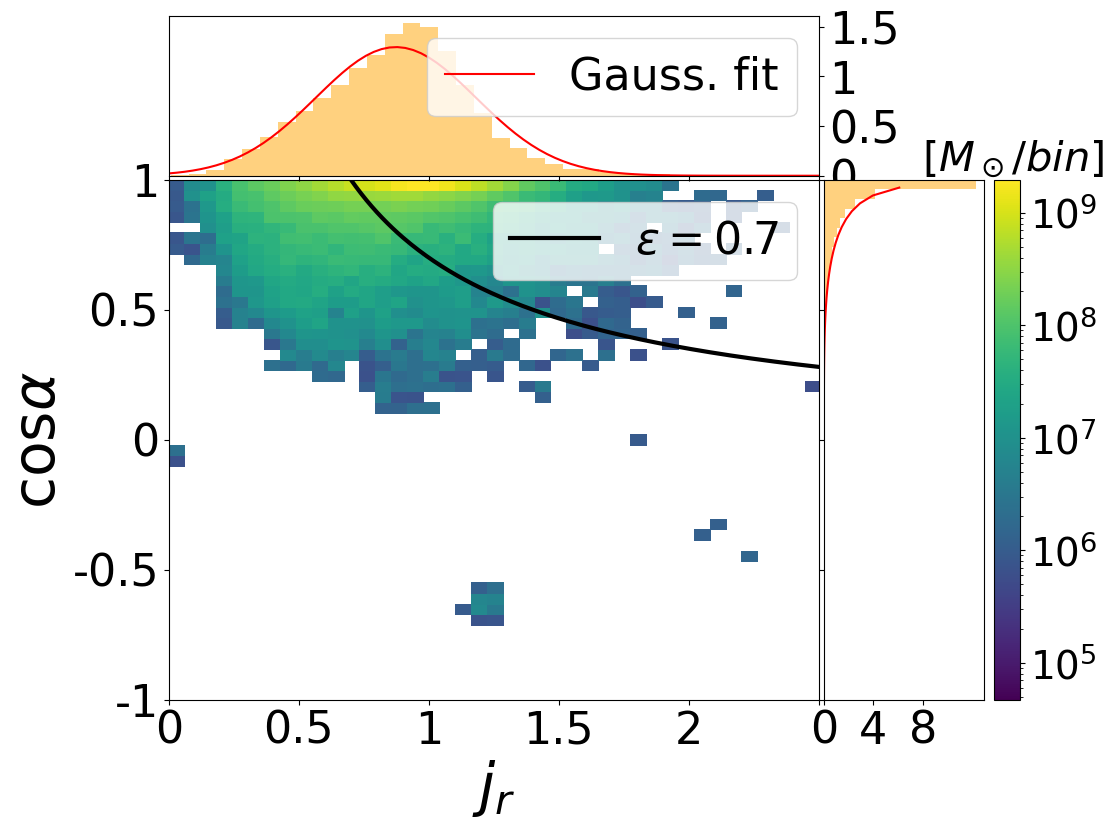}
\caption{\label{mc3_c}The disc component after Monte Carlo; note concentration at $\cos\alpha \sim 1$, and $j_{r}\sim1$}
\end{subfigure}
\caption{\label{mc3}Mass density distribution  
of a typical two-component disc-dominated galaxy of $\log_{10}(M_*/M_\odot) \sim 11$   
in the $\cos\alpha$ versus  $j_{r}$  plane, the colorbars are in units of $M_\odot$ per bin with arbitrary normalization.  The black line is the $\varepsilon = 0.7 $ cut that is widely used in the literature to separate disc stars from bulge stars. Since the black line cuts the locus of disc particles in half, it underestimates the disc fraction, and in the process contaminates both populations.  
The middle and right plots show the bulge and disc components, respectively, after Monte Carlo separation (described in \S~\ref{2d-method})  
The red curves are the maximum likelihood fit to the 1D marginal distributions.   
In b) we fit the marginal distribution using a Gamma distribution for $j_\text{r}$ and uniform for $\cos\alpha$.
 For 
 c), the correlation coefficient between the variables is  $+0.1$.   
 We fit the 1D marginal distributions using a Gaussian distribution for  $j_{r}$  
 and a Beta distribution for $\cos\alpha$. Together with the copula (Eq.~\ref{p_disc_eq})  quantifying correlations in this 2D space, these distributions fully specify the distribution of stellar particles in the disc.  
 }
\end{figure*}

\begin{figure*}
\begin{center}
$\begin{array}{c@{\hspace{0.5in}}c}
\includegraphics[width=6.6in,angle=0]{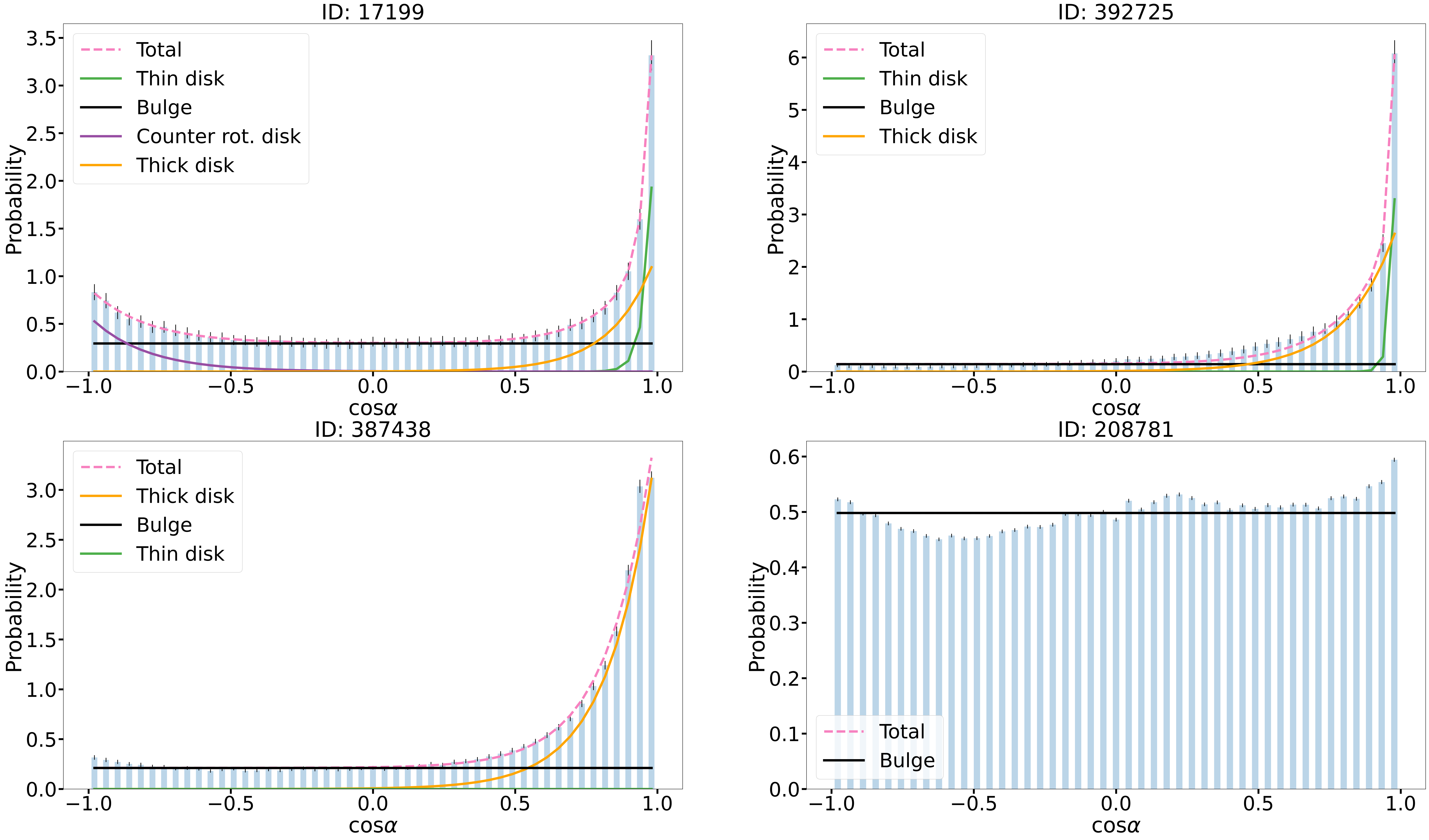}  
\end{array}$
\caption{ \label{1dcos} The 1-dimensional distributions for $\cos{\alpha}$ for four galaxies exhibiting different levels of substructure at approximately equal stellar mass, $  \log_{10}(M_*/M_\odot) \sim 11$. 
The ID numbers given in the title for each panel are the subhalo IDs from TNG100.
Galaxy-17199 exhibits four substructures; Galaxy-392725  
statistically favors a fit with a bulge and two disc components, whereas Galaxy-387438 favors a bulge with a single disc; finally, Galaxy-208781 appears to be a dispersion-dominated elliptical galaxy.  
The red curve is the least-squares fit to the histogram using Eq.~\eqref{1d_eq}   
Also, the contribution from each term is plotted separately. 
}
\end{center}
\end{figure*}

The $\varepsilon$ parameter has been widely used in literature \citep[e.g.,][]{Abadi_2003}; typically stars with $\varepsilon > 0.7$ are identified as disc stars, and the rest as bulge stars. 
Applying this method to IllustrisTNG, we calculate the fraction of disc stars $f^\text{disc}_\mathrm{\varepsilon}$ 
in each galaxy and identify galaxies with $f^\text{disc}_\mathrm{\varepsilon} > 0.3$ 
as disc-dominated galaxies.  
As shown in the left panel of Fig.~\ref{mc3}, which illustrates this methodology for one individual galaxy (a typical disc-dominated galaxy of medium mass), decomposition into components based on the $\varepsilon$ parameter provides results that are not physically meaningful. The 2D histogram of the star particle in the $\cos\alpha$ versus $j_\text{r}$ plane shows that  there is a locus of star particles that clearly corresponds to the disc component (near $\cos\alpha=1$).  However, the curve with $\varepsilon=0.7$ cuts this locus into two parts, omitting some disc particles from the disc and therefore underestimating the disc fraction significantly. At the same time, it includes a significant part of the 2D plane that does not appear to be part of the disc, thereby contaminating the disc with bulge stars. As shall be seen in \S~\ref{n_frac} this effect is pervasive within the galaxy sample, rather than being a feature of this specific galaxy.  
Based on the discussion in \S~\ref{kin_params} about disc star particles' behavior in kinematic phase space,  as a first step to 
estimate the fraction of disc stars, we consider a simple 1D model based on the $\cos\alpha$ parameter defined in \S~\ref{kin_params}. 
This model compared to GMM, k-means and other multi-parameter methods is easy to interpret, as it is based on a clear single dynamical parameter; and compared to $\varepsilon$ it discriminates more effectively between the components.  
For disc-dominated galaxies, we expect to see a population that sharply rises towards $\cos\alpha = 1$ identified as the thin disc, a second population that slowly rises towards $\cos\alpha = 1$ identified as the thick disc, a
third population with $\cos\alpha \sim -1$  identified as a counter-rotating disc, and a uniformly distributed population identified as the bulge. Hence we consider a model of the form:
\begin{equation}
 p(\cos\alpha) =  \underbrace{A \cdot e^{B\cos\alpha} }_\text{thin disc}+ \underbrace{D \cdot e^{F\cos\alpha}}_\text{thick disc} + \underbrace{G \cdot e^{-H\cos\alpha}}_\text{counter-rotating disc} +  \underbrace{C}_\text{bulge}
\end{equation}
where $A, B, C, D, F, G, H > 0$ and $B > F$. Since disc stars appear to ``grow'' towards $\cos\alpha =1$, we 
choose the exponential function on an empirical basis. We fit  the model to the data using the maximum likelihood  method, using initial parameter guesses from least squares fit to the histograms.  
An illustration of this model is given in Fig.~\ref{1dcos} for four different galaxies at $\log_{10}(M_*/M_\odot) \sim 10 $. The first galaxy has noticeable populations in all four components mentioned above, the second galaxy lacks the counter-rotating disc, the third galaxy exhibits a statistical preference for only one disc component, and the last galaxy appears to be dominated by a bulge component.  Simply by considering the direction of the stars' movement with respect to the overall galaxy angular momentum, we can identify as many as four components.  Further, the statistical robustness of this method will be explored in \S~\ref{thin-thick-section}.  
Using the best-fitting models, we can define the mass fraction of each component ($f^\text{thin}_\mathrm{\cos\alpha}$, $f^\text{thick}_\mathrm{\cos\alpha}$, $f^\text{cr}_\mathrm{\cos\alpha}$, $f^\text{bulge}_\mathrm{\cos\alpha}$) by integrating the corresponding term.  We also define  $f^\text{disc}_\mathrm{\cos\alpha} = f^\text{thin}_\mathrm{\cos\alpha} + f^\text{thick}_\mathrm{\cos\alpha} + f^\text{cr}_\mathrm{\cos\alpha}$ as the total mass fraction of the stars in the disc. 
In \S~\ref{results},  
we discuss the ensemble statistics of the population based on use of this 1D decomposition model.

\subsection{Model discrimination using Akaike Information Criterion}
\label{AIC}

For the 1D model, in order to discriminate galaxies with two disc components from one disc component, and to identify galaxies with a  counter-rotating disc component, we employ the Akaike information criterion (AIC) for selecting the model. The AIC was
developed by \cite{Akaike}  and is widely used in statistical astrophysics. The AIC is defined
\begin{equation}
{\rm AIC} = -2  \cdot {\rm ln} (\widehat{L}) + 2k  , 
\end{equation}
\noindent
where ${\rm ln} (\widehat{L})$ is the log-likelihood of the data set evaluated at the best-fitting model parameters, and
 $k$ is the number of free parameters 
in the model. The second term is a penalty for the number of 
parameters introduced in the fitting function, which helps avoid overfitting. For a given galaxy, the model with the smaller AIC is chosen. Thus, AIC basically measures how far a model is from the truth.
However, if the AIC values of two models are very close, then choosing the statistically preferred model becomes hard. Therefore, to account for the possibility of selecting the less preferred model, a threshold value is adopted, so that models that are within this threshold are considered to be of equal rank. We use a threshold value of 10 for the difference in AIC values   
to discriminate between  models, in order to ensure the selected model is truly the statistically preferred one \citep{liddle}. 
 
\subsection{2D kinematic decomposition model } \label{2d-method}
 
In this subsection we extend the previous model by considering an additional parameter in order to achieve more accurate results as discussed in \S~\ref{kin_params}. When applying this model, we choose to probabilistically assign each particle to a galaxy component. However, when doing so, this 2D model identifies two components, compared to the 1D model, which can identify as many as four components.    
The model identifies components through two physically-motivated assumptions:
\begin{enumerate}
  \item Disc stars' angular momentum is aligned with the total angular momentum of the galaxy, while the orientation of  bulge stars' angular momentum is random.
  \item Disc stars' orbits are approximately circular, while bulge star's orbits are elliptical.
\end{enumerate}
Assumption (i) is captured by use of the $\cos\alpha$ parameter in the model.  For assumption (ii), we adopt the parameter $j_\text{r}$ as described in \S~\ref{kin_params}.  
 To build this 2D model, we consider Fig.~\ref{corner} subfig.~3, which shows a 2D histogram of star particles in the $\cos\alpha$ versus $j_\text{r}$ plane for a single galaxy.  There is a semi-ellipse-shaped population at $\cos\alpha \sim 1$, and another population that stretches the whole interval of $\cos\alpha$.  We assume bulge stars should exhibit a flat distribution in  $\cos(\alpha)$, so their distribution in this plane should be solely dependent on  $j_{r}$.  
 On the other hand, we assume the distribution of disc stars depends on both parameters, consistent with what is shown in this figure. 
Hence, we consider the following model for the probability distribution of star particles in the $(j_\text{r}, \cos\alpha)$ space
\begin{equation}\label{1d_eq}
   p_\text{star}(j_\text{r},\cos\alpha) \equiv (1-f^\text{disc}) \, p_\text{bulge} (j_\text{r},\cos\alpha) + f^\text{disc} \, p_\text{disc}  (j_\text{r} , \cos\alpha)
\end{equation}
where $p_\text{bulge}$ and $p_\text{disc}$ are the normalized probability that a given star belongs to the bulge or the disc.   
In Fig.~\ref{ellip} we show three bulge-dominated galaxies of various mass. 
These three galaxies illustrate the universal signatures of bulge-dominated galaxies. As expected, the star particles are found to be uniformly distributed in $\cos\alpha$ and to mostly follow an elliptically shaped orbits as evidenced by the skewed distribution of $j_\text{r}$  with a peak around  $j_\text{r} = 0.5$.
Based upon the aforementioned signature, we build our model for the bulge component.
Also, our assumptions regarding the bulge component lead to an assumption of independent distributions for the two parameters:
\begin{equation}
p_\text{bulge} (j_\text{r},\cos\alpha) = \Gamma (j_\text{r}) \cdot U(\cos\alpha) 
\end{equation}
where $\Gamma(j_\text{r})$ is the Gamma  distribution and $U(\cos\alpha)$ is the Uniform distribution on the interval [-1,1]. The Gamma   
distribution was chosen empirically to fit the skewed distribution in $j_\text{r}$. 
For the distribution of the star particles in the disc we will use non-parametric representation via kernel density estimation to infer $p_\text{disc}$. 

\begin{figure}
 \includegraphics[width=\columnwidth]{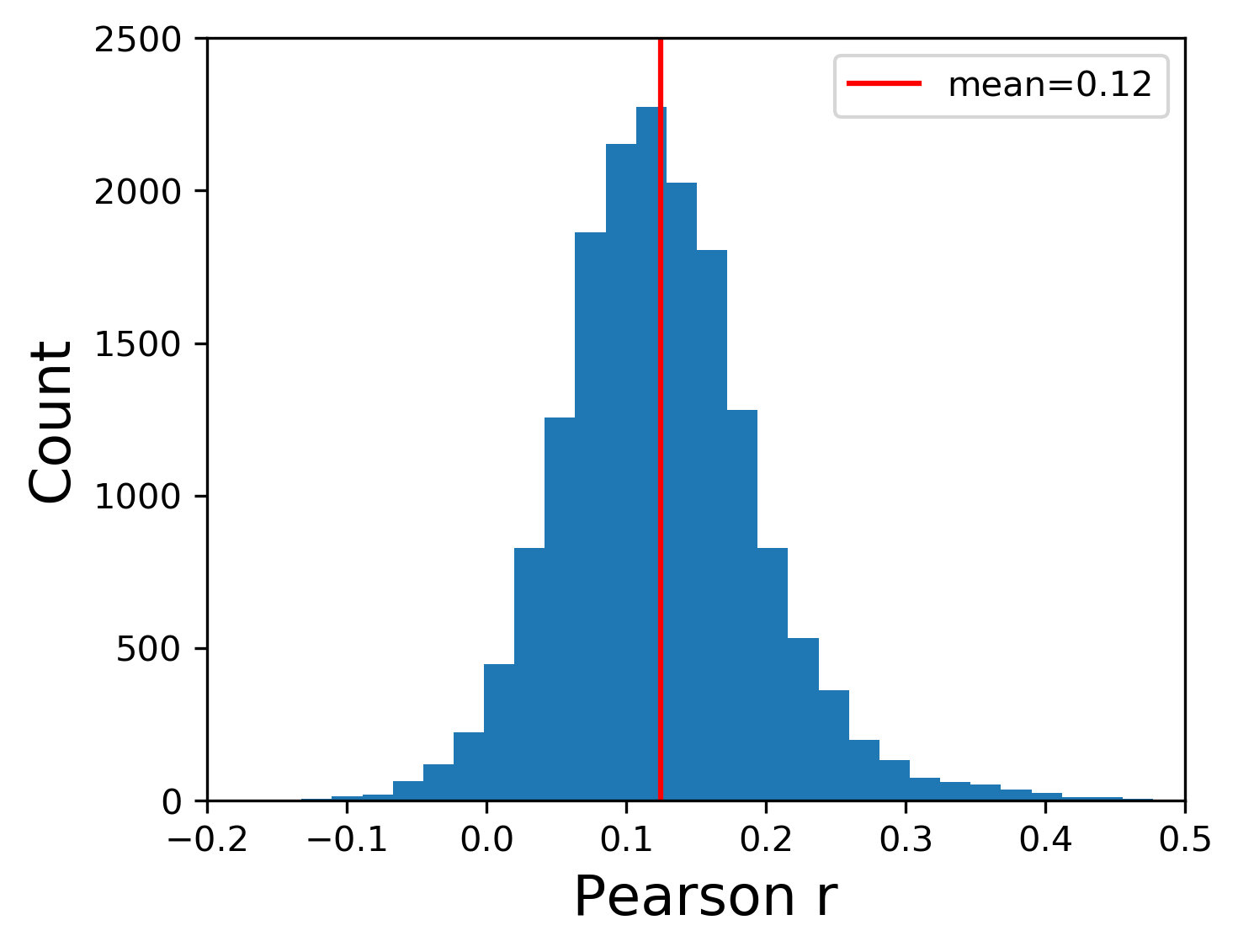}
 \caption{\label{pearson}
 Distribution of Pearson $r$ correlation coefficients for $j_\text{r}$ and $\cos\alpha$ parameters in the disc component (Fig.~\ref{mc3}c) for the whole galaxy ensemble. This correlation is captured by the \textit{copula} structure. Since the correlation is weak for the whole galaxy population at $\sim +0.1$. 
 }

\end{figure}

 It is useful for the purpose of plotting and for future studies in identifying additional structures to derive a parametric form of the joint distribution $p_\text{disc}(j_\text{r},\cos\alpha)$ to complement the non-parametric model. We therefore model the marginal distributions of $p_{\rm disc}(j_\text{r})$ and $p_{\rm disc}(\cos\alpha)$ individually using a Gamma distribution for the former and a uniform distribution for the latter. The fitting is done using maximum likelihood estimation. In order to form a joint distribution $p_\text{disc}(j_\text{r},\cos\alpha)$ using the fitted models for the marginals, we use a copula model \citep{Skla59}
\begin{equation}\label{p_disc_eq}
 p_\text{disc}  (j_\text{r} , \cos\alpha) = c(P_\text{disc}(j_\text{r}), P_\text{disc}(\cos\alpha) ) \cdot p_\text{disc}(j_\text{r}) \cdot p_\text{disc}(\cos\alpha) \, .
\end{equation}
Here $c(P_\text{disc}(j_\text{r}), P_\text{disc}(\cos\alpha) )$ is the cupola density and $P_\text{disc}(j_\text{r})$ and $P_\text{disc}(\cos\alpha)$ are the cumulative distribution functions of the aforementioned marginals $p_{\rm disc}(j_\text{r})$ and $p_{\rm disc}(\cos\alpha)$. The cupola density is used to model the correlation between two random variables to form a joint distribution. In this work we assume a Gaussian correlation structure and therefore use a Gaussian cupola as defined in \citet{gauss-copula1} and~\citet{gauss-copula2}. We refer the reader to the literature for more details on the cupola method. Besides the cumulative distribution function of the marginals, the cupola density depends on the correlation matrix between $j_\text{r}$ and $\cos\alpha$. We estimate this with the Pearson $r$ correlation coefficient.  For our case, as seen in Fig.~\ref{pearson}, the correlation coefficient 
is low with a mean value of $\sim$+0.1 for the $j_\text{r}$ and $\cos\alpha$ parameters in the disc structure for the whole sample.

Now we list the steps necessary to build the deterministic 2D model for each galaxy:
 
\begin{enumerate}
    \item To determine  $p_\text{bulge} (j_\text{r},\cos\alpha)$, we need a sample of  particles that has a flat distribution in $\cos\alpha$. To do so, based on the 1D model, we make a cut on the $\cos\alpha$ parameter to use only the particles with $\cos\alpha$ values that are dominated by the bulge component.  The cut is carried out in an automated way based on identifying the value of $\cos\alpha$ at which the exponential term (representing the disk) has a value equal to 10\% of the flat term (representing the bulge)\footnote{One caveat of making such a cut is that sometimes too few particles are left after the cut. In our sample, about 10 per cent of the galaxies suffer from this issue, with fewer than 500 particles after the cut. However, all of these galaxies have $\log(M_*/M_\odot)<10$ and in \S~\ref{n_frac} our exploration of the entire galaxy sample will motivate us to exclude these low-mass galaxies entirely from our analysis.  
    }.    
    
    \item We then fit a Gamma distribution to the $j_\text{r}$ distribution for these cut-out particles, using maximum likelihood estimation. After doing so, we have the full explicit form for $p_\text{bulge}(j_\text{r},\cos\alpha)$.   
    \item  $1-f^\text{disc}$ can be estimated   
    from the 1D model, i.e. 
    $f^\text{disc} = f^\text{disc}_\mathrm{\cos\alpha}$, after which we have the full bulge term 
    $(1-f^\text{disc}_\mathrm{\cos\alpha} ) p_\text{bulge} (j_\text{r},\cos\alpha)$.  
    \item Next, in order to determine the total probability $(1-f^\text{disc}_\mathrm{\cos\alpha}) \, p_\text{bulge} (j_\text{r},\cos\alpha) + f^\text{disc}_\mathrm{\cos\alpha} \, p_\text{disc}  (j_\text{r} , \cos\alpha)$, we need to estimate the joint probability distribution of the 2D plane for all particles. We already determined the first term in step (iii). The second term with  $p_\text{disc}$  
    can be estimated non-parametrically by kernel density estimation.  
    \item Thus, we have the probability that a star particle is in the bulge based on its $j_\text{r}$ and $\cos\alpha$ values: 
    \begin{equation}\label{eq:bulgep}
       \eta_{\rm bulge}= \frac{(1-f^\text{disc}_\mathrm{\cos\alpha})  p_\text{bulge} (j_\text{r},\cos\alpha)}{ (1-f^\text{disc}_\mathrm{\cos\alpha}) \, p_\text{bulge} (j_\text{r},\cos\alpha) + f^\text{disc}_\mathrm{\cos\alpha} \, p_\text{disc}  (j_\text{r} , \cos\alpha)}
    \end{equation}
    
\end{enumerate}
    
    Upon building the 2D model deterministically, we can generate Monte Carlo realizations of the model, assigning star particles to the bulge by comparing $\eta_{\rm bulge}$ from  Eq.~\eqref{eq:bulgep} (evaluated at each star particle's value of $j_\text{r}$ and $\cos\alpha$) to uniformly-distributed floats in the range $[0,1]$. The remaining particles are identified as belonging to the disc.  
A single realization of this simulation for a single galaxy is shown in Fig.~\ref{mc3}.  
Recall that  Fig.~\ref{mc3} has the same axis as Fig.~\ref{corner} subfig.~3;   
it shows all the particles before the Monte Carlo simulation, in the form of a mass density distribution in the $\cos\alpha$ versus $j_\text{r}$ plane.  The subplots~b) and ~c) show the bulge and disc components, respectively, after the Monte Carlo separation process described above.  

In Fig.~\ref{mc3_b}, the bulge component isolated by the Monte Carlo 
method has the same structure in the $\cos\alpha$ versus $j_\text{r}$ plane as elliptical galaxies in Fig.~\ref{ellip} by construction.  
Similarly, in Fig.~\ref{mc3_c},  
the Monte Carlo simulation picks out the locus of particles near $\cos\alpha \sim$ 1 and $j_\text{r} \sim$ 1, also by construction. For the disc component of this galaxy, the correlation coefficient between $\cos\alpha$  and $j_\text{r}$  is $+0.1$, as measured by the Pearson $r$ coefficient. As discussed earlier in this section, the correlation is captured by the copula; hence we have the full explicit functional form for the stellar particles in the disc. Also, both Figs.~\ref{mc3_b} and~\ref{mc3_c} show how using a  cutoff on the circularity parameter (denoted by the black curve) contaminates the two components, whilst undercutting the disc component. As a quantitative illustration of this claim, the disc fraction  values of the different methods for the specific galaxy in Fig.~\ref{mc3} are 
$ f^\text{disc}_\mathrm{\varepsilon}  =0.44$,  $f^\text{disc}_\mathrm{\cos\alpha}  = 0.72$,  $f^\text{disc}_\mathrm{2D}  = 0.74$.

 We calculate the fraction of disc stars $f^\text{disc}_\text{2D}$   
 in each galaxy by summing over the particles assigned to the disc population in a single realization, and compare it with other methods. We therefore have three different  $f^\text{disc}_{x}$ methods, where $x=\varepsilon, \, \cos\alpha, \, \text{2D} $. However, one should keep in mind that by construction the expected value for $f^\text{disc}_{\cos\alpha}$ is $f^\text{disc}_\text{2D}$, assuming model sufficiency.
 
 Statistical uncertainties on ensemble quantities such as number fraction of disk-dominated galaxies at a given stellar mass are expected to be very small because uncertainties on a per-galaxy basis scale with $1\big/\sqrt{N}$. 

 Also, systematic uncertainties due to the baryonic physics implementation, and statistical uncertainty due to cosmic variance  should be much higher.
On the other hand, statistical uncertainties on measured quantities for individual galaxies  will demonstrate the probabilistic nature of the 2D model, and could potentially help us identify regions of parameter space  where   measurements of individual galaxy properties maybe not be reliable. We demonstrate and investigate the probabilistic elements of the model in \S~\ref{rsd_sec}.


\begin{figure*}\label{ellip}
\begin{subfigure}{.3\textwidth}
\centering
\includegraphics[width=5.5cm ]{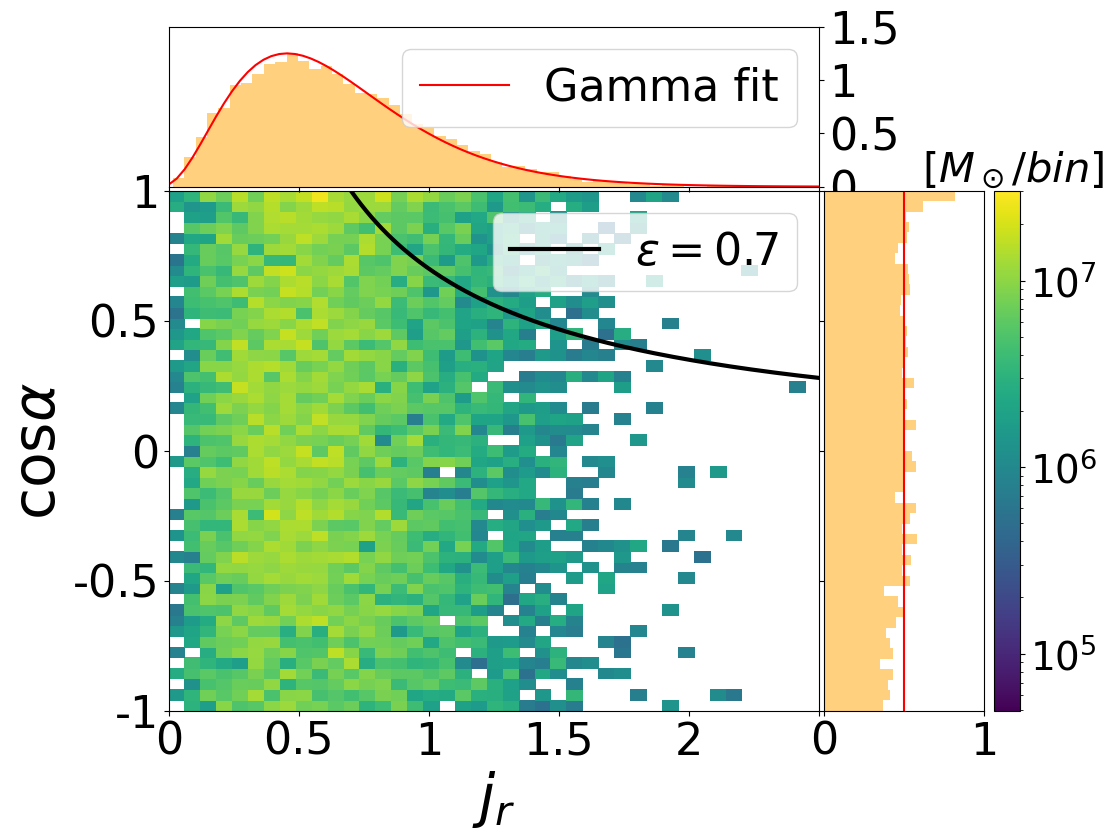}
\caption{$\log_{10}(M_*/M_\odot) = 10  $}
\end{subfigure}\hfill
\begin{subfigure}{.3\textwidth}
\centering
\includegraphics[width=5.5cm ]{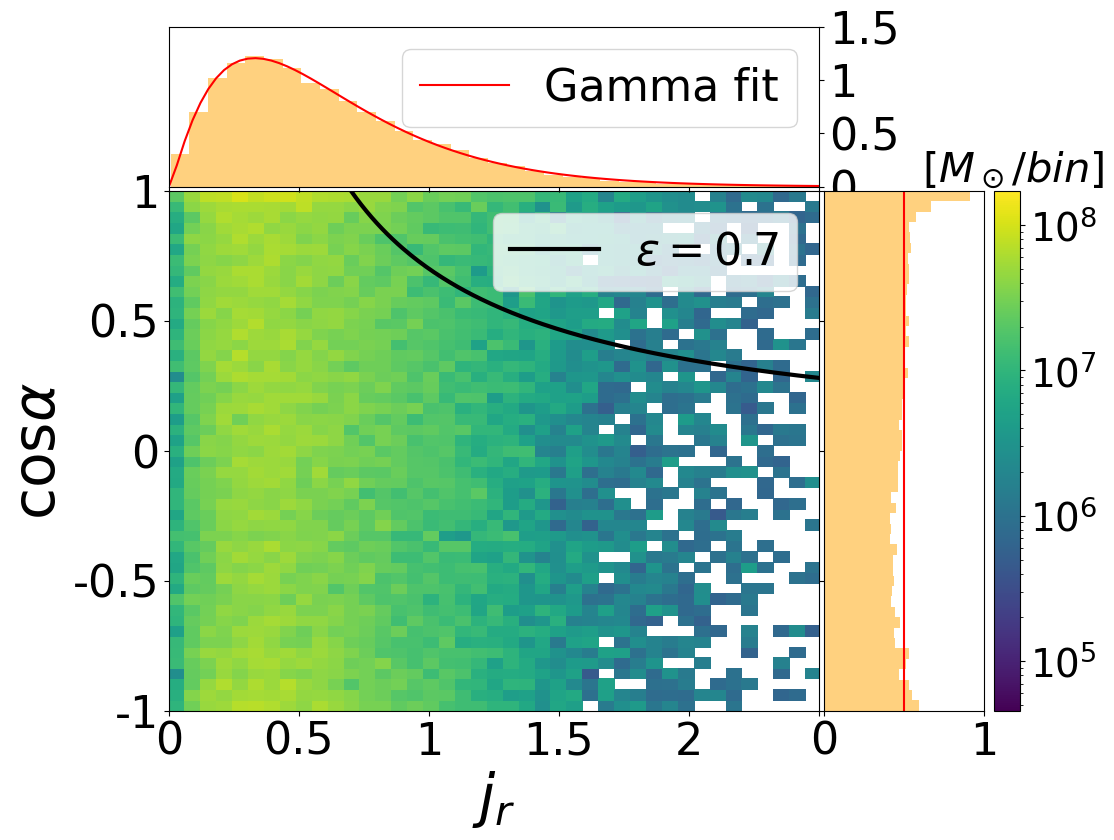}
\caption{ $\log_{10}(M_*/M_\odot)=10.5  $}
\end{subfigure}\hfill
\begin{subfigure}{.3\textwidth}
\centering
\includegraphics[width=5.5cm ]{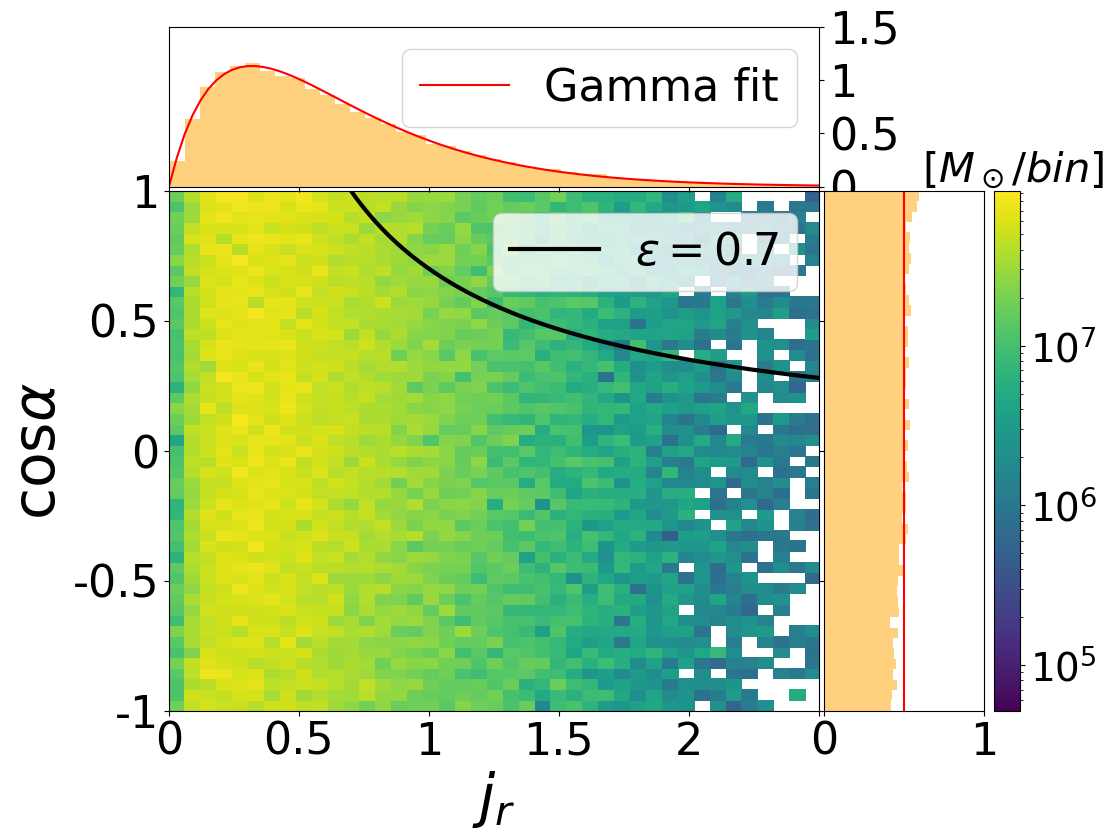}
\caption{$\log_{10}(M_*/M_\odot)=12  $}
\end{subfigure}
\caption{\label{ellip}  
Mass density distribution of three elliptical galaxies with different masses in the $\cos\alpha$ versus  $j_{r}$  plane. The three panels show all star particles in the galaxy, like Fig.~\ref{mc3_a}.   
Since $\cos\alpha$ measures alignment with the total angular momentum, even for a dispersion-dominated galaxy, at least a small fraction of the star particles will be distributed closer to  $\cos\alpha \sim 1$ degrees; that is also why there is a mild deficit at negative $\cos\alpha$, which is most easily visible in the 1D histograms on the side of each panel.  The mostly flat distribution in $\cos\alpha$ and the skewed distribution in $j_r$   
are universal signatures for elliptical galaxies (excluding tidally disturbed and irregular galaxies).  
}
\end{figure*}

\section{Results - Ensemble statistics}\label{results}


Here we investigate the ensemble statistics derived using the 1D and the 2D kinematic decomposition models.  
First, we compare number fraction of disc-dominated galaxies at a given stellar mass bin derived from the 1D and 2D models with observational data. 
Second, we study the distribution of $f^\text{disc}_x$ for the whole ensemble,  
and its dependence on mass  and color. Third, we investigate the contribution from the decomposed components to the total galaxy mass budget and compare with observed values. Fourth, using a 1D radial surface density profile of classified/decomposed galaxies we validate the model against observation. Lastly, we explore the probabilistic nature of the 2D model and present uncertainties on the measured quantities.  

\subsection{The number fraction $N_\text{disc}$ versus stellar mass } \label{n_frac}
 
\begin{figure} 
 \includegraphics[width=\columnwidth]{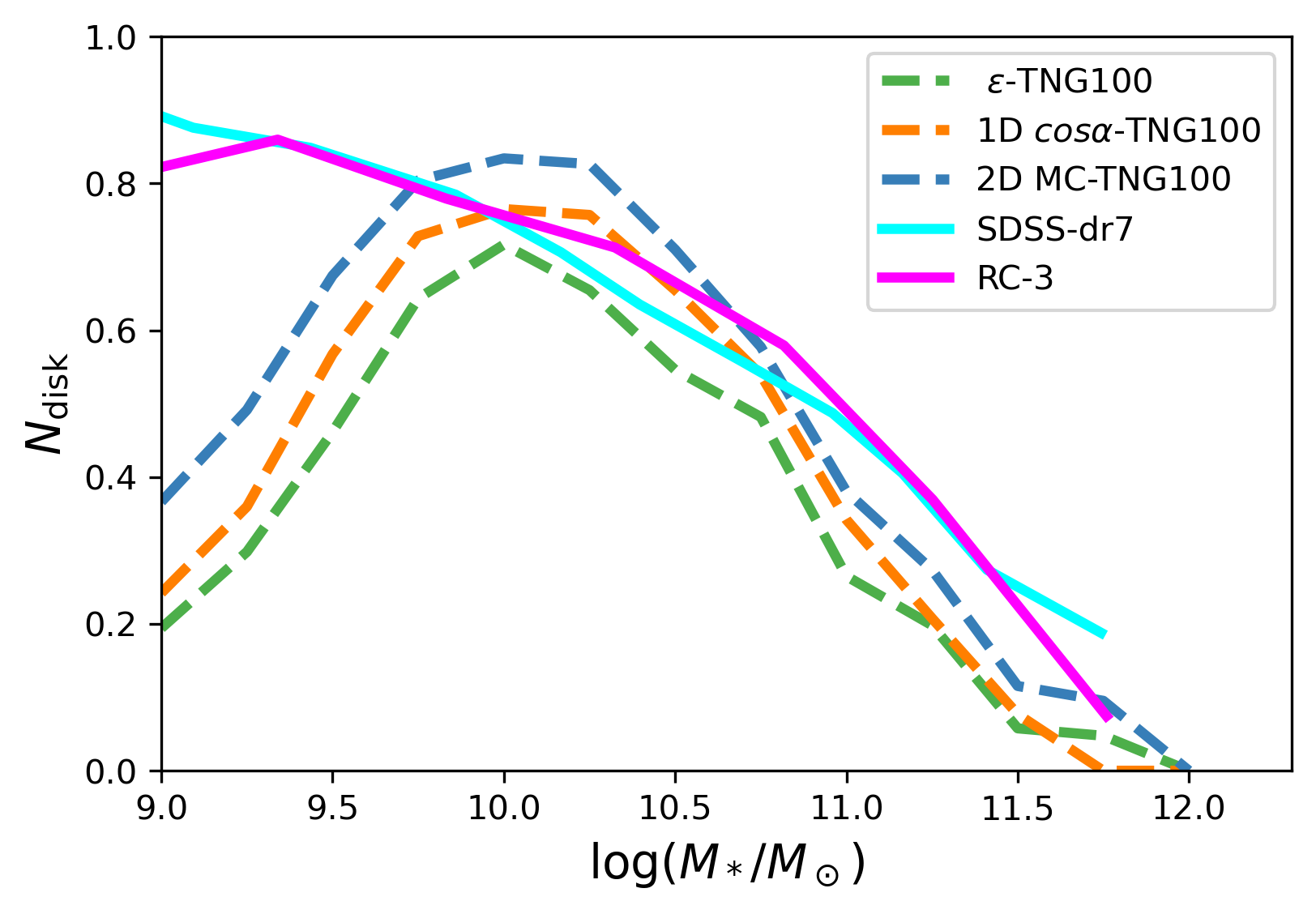}
 \caption{\label{f_disc vs mass}Number fraction of disc-dominated galaxies as a function of stellar mass. The dashed lines show the results for IllustrisTNG100 using the three different kinematic decomposition methods presented in the text, while the solid lines are from observations. The agreement is good at high mass ($ \log_{10}(M_*/M_\odot) > 10$),
 but for the IllustrisTNG100-1 at low mass, the fraction decreases, which may be attributed to the resolution limits of the simulation. In the following sections we will impose a $ \log_{10}(M_*/M_\odot) > 10$ mass cut, indicated by the black dotted line.  
 }

\end{figure}

Using the three kinematic decomposition methods described in \S~\ref{methods}, we can now calculate $N_\text{disc}$, the number fraction of disc-dominated galaxies  
at a given stellar mass.   
The cutoff used for $f_\varepsilon^\text{disc}$ to classify disc-dominated galaxies varies from study to study; here we chose  $f_\varepsilon^\text{disc} > 0.3$, or a bulge-to-total ratio (BTR) $< 0.7$ for disc-dominated galaxies.  
For the other two methods, as will be explored in more detail in the following section, we define disc-dominated galaxies as those with   bulge-to-total ratio (BTR) $<0.5$.  
In Fig.~\ref{f_disc vs mass} we show the three curves for $N_\text{disc}$ versus stellar mass, along with observational data from \cite{bluck-sdss}, which uses r-band data from the SDSS data release 7 (SDSS-DR7), and \cite{conselice}, which uses data from Reference Catalog 3  \citep[RC-3;][]{rc3}. \citet{conselice} generated a nearly volume limited sample from RC-3 by employing absolute magnitude and redshift cuts on the galaxies. The SDSS-DR7 sample yields statistics that are expected for a volume limited sample by applying weights based on the inverse of the volume over which any given
galaxy would be visible in the survey.  Imposing a lower mass cut of $ \log_{10}(M_*/M_\odot)=9 $ on all samples enables a reasonable comparison to IllustrisTNG100 data.  
Also, the SDSS-DR7 data allows a direct comparison to the results of our 1D and 2D models, since \cite{bluck-sdss} used the same BTR$<0.5$ threshold, but based on a decomposition of light profiles  
to define the sample of disc-dominated galaxies. On the other hand, disc-dominated galaxies in the RC-3 catalog were identified by eye, which provides a more approximate but independent method.

From this panel we see that $\varepsilon$ underestimates the disc fraction compared to the observational data and  the other methods, despite the fact that the BTR<0.7 cut in principle should identify more disc-dominated galaxies, so the fact that it is below the other methods is indicative of a problem. Nevertheless, it exhibits a similar trend with mass as the other methods.  
The two methods of $\cos\alpha$ and 2D Monte Carlo provide very similar results, by construction. At high mass, $\log_{10}(M_*/M_\odot)>11  $, the difference is about 5\%, but for intermediate mass of $\log_{10}(M_*/M_\odot) \sim 10.25  $ and lower, the 2D methods is higher by about 20\%.  
Conversely the observational data from SDSS-DR7 is slightly higher than the 1D and the 2D methods  %
by 10\% at high mass, and then at $\log_{10}(M_*/M_\odot) \sim 10.7  $ meets the curves from TNG100.
Nonetheless, at low mass, $\log_{10}(M_*/M_\odot) < 9.7$, the curves start to diverge, with the curves from IllustrisTNG100-1 declining significantly to roughly 0.4 at $\log_{10}(M_*/M_\odot) = 9  $, while the curves from observational data reach 0.9. For this reason we suspect that at masses lower than $ \log_{10}(M_*/M_\odot)=9.5 $, the resolution effects may start to suppress the number of disc-dominated galaxies. Supporting evidence and investigation into this issue has been presented in \cite{tacchella2019}.  
As stated in \cite{tacchella2019}, the resolution of the simulations may have an effect on the exact values of the spheroid-to-total ratio (which is $1-f^\text{disc}$),  
but the overall trends with stellar mass are robust. Further, they state that the growth in time of disc fractions of low-mass galaxies is underestimated 
in the simulations.  
Hence, we conclude that the curves for disc fractions from our bulge-disc decomposition method as applied to TNG100-1 agree well with observational values down to $\log_{10}(M_*/M_\odot) = 10$, below which the resolution of the simulation may be affecting the results. We therefore restrict ourselves to $\log_{10}(M_*/M_\odot) \ge 10$ in the rest of the paper.  
\subsection{Disc fraction dependence on mass and color}\label{mass-color}

\begin{figure}
\begin{subfigure}{.3\textwidth}
\centering
\includegraphics[width=7cm]{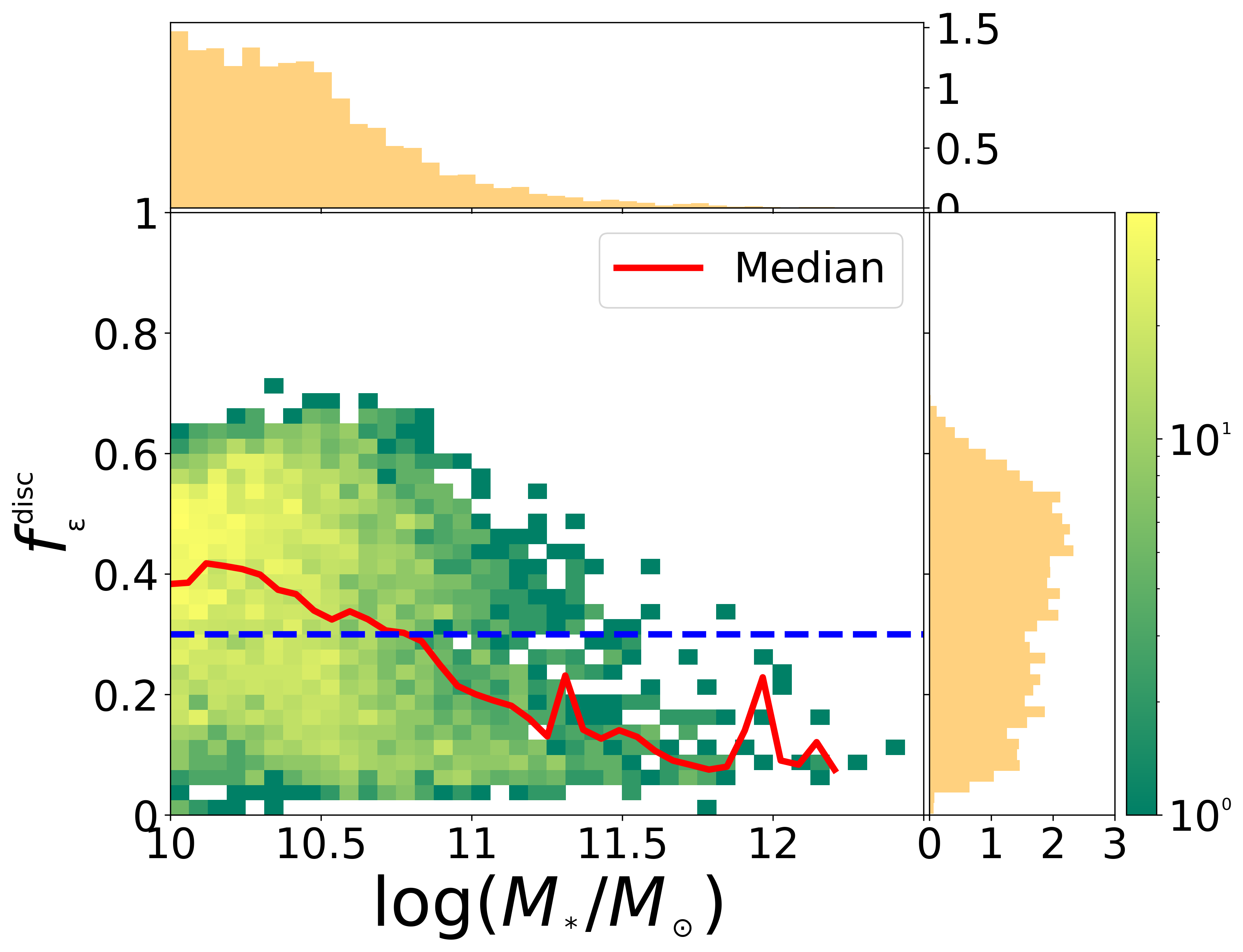}
\end{subfigure}\hfill
\begin{subfigure}{.3\textwidth}
\centering
\includegraphics[width=7cm]{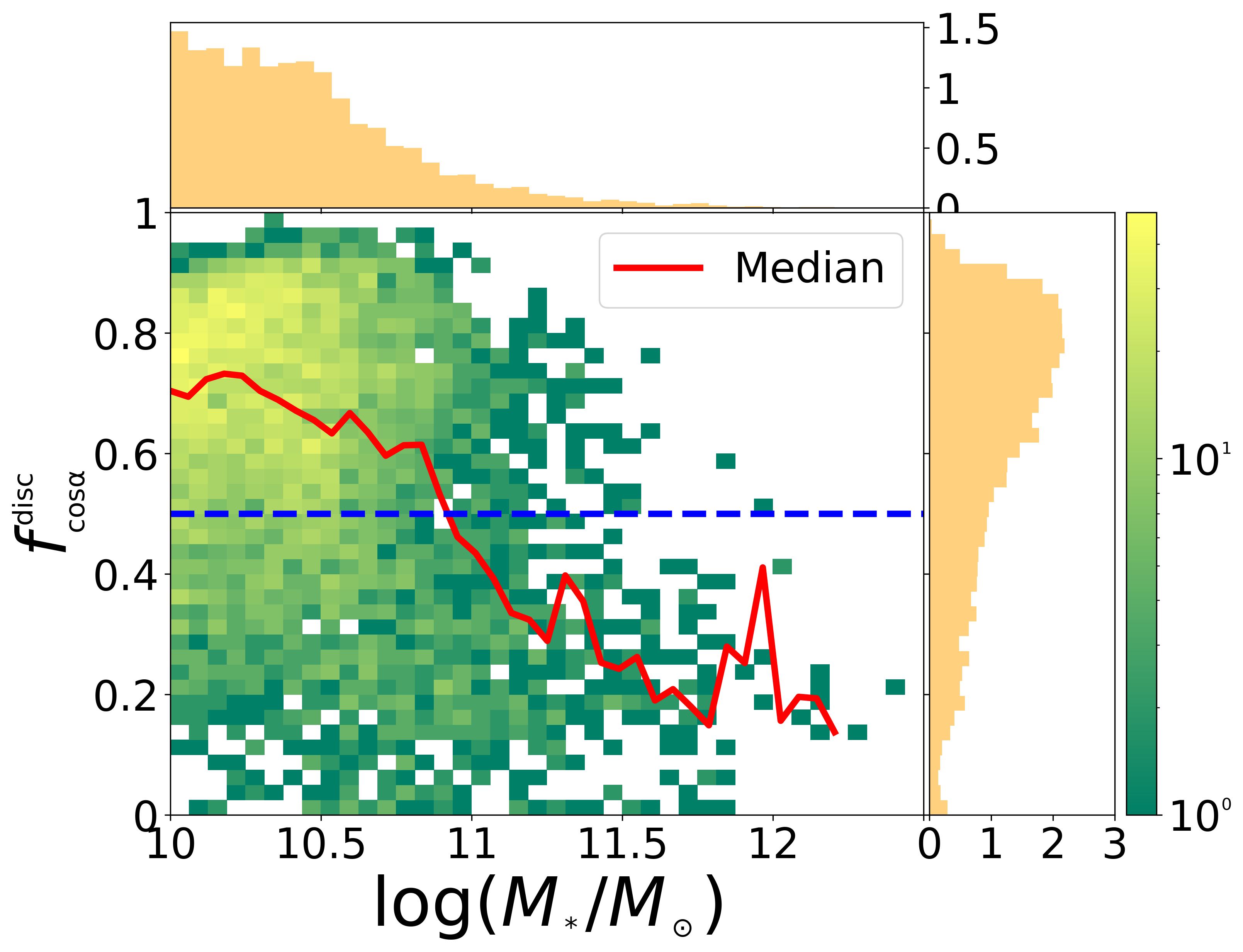}
\end{subfigure}\hfill
\begin{subfigure}{.3\textwidth}
\centering
\includegraphics[width=7cm]{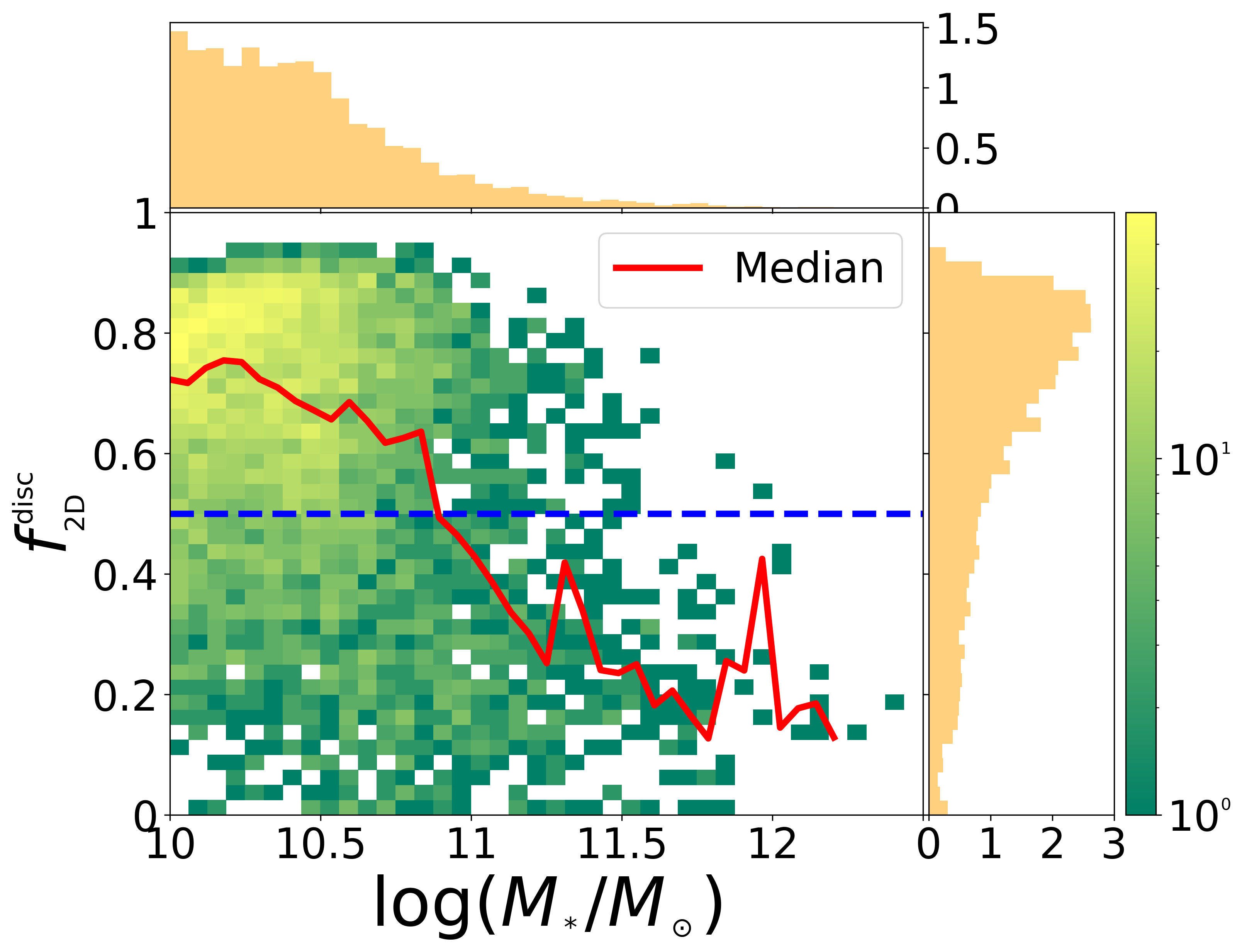}
\end{subfigure}
\caption{\label{fstar_vs_mass} The top, middle, and bottom panels show the 2D planes of  $f^\text{disc}_\mathrm{\varepsilon}$, $f^\text{disc}_\mathrm{\cos\alpha}$ and $f^\text{disc}_\text{2D}$ versus stellar mass. 
There is no obvious bimodality in $f^\text{disc}_x$ at fixed mass in any of the methods. $f^\text{disc}_\mathrm{\varepsilon}$ typically produces lower disc fractions than the other two methods; we have previously demonstrated that this is due to its exclusion of some disc particles from the disc (Fig.~\ref{mc3_a}). The 2D histograms of $f^\text{disc}_\mathrm{\cos\alpha}$  and $f^\text{disc}_\text{2D}$ versus stellar mass appear very similar to each other due to the fact that the $f^\text{disc}_\mathrm{\cos\alpha}$ and $f^\text{disc}_\text{2D}$ are highly correlated.   
The colorbar indicates the number of galaxies in each bin. The blue dashed line denotes the cut for disc dominated galaxies; for details please refer to \S~\ref{mass-color}. 
The red curve corresponds to the median $f^\text{disc}_x$ in a given stellar mass bin. }   
\end{figure}

\begin{figure}
 \includegraphics[width=\columnwidth]{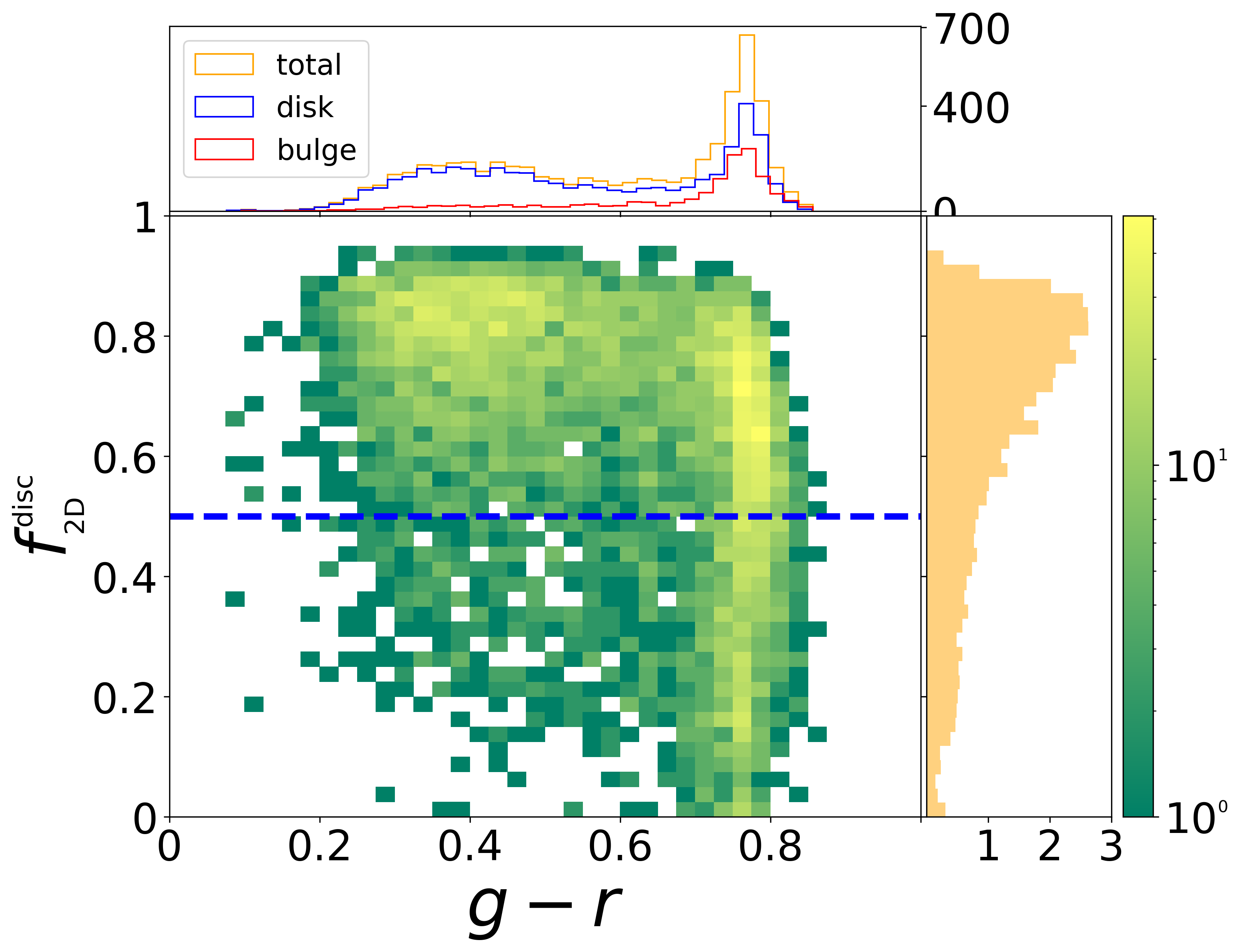}
 \caption{\label{g-r-color} Rest-frame color $g-r$ as a function of stellar disc fraction.  
 Blue dashed line at 0.5 is the cutoff for disc-dominated galaxies. The 
 majority of the galaxies here are blue disc-dominated galaxies  
 centered around $g-r \approx 0.4$. There is a significant number of red disc 
 galaxies at $g-r \approx 0.7$. This excess amount corresponds to 
 IllustrisTNG100 producing red galaxies of intermediate mass 
 ($\log_{10}(M_*/M_\odot)=10.5 - 11.5 $ 
 ) more than expected \protect\citep{tng-bimodal}.  
 }
 \label{fig:f_disc vs mass}
\end{figure}

In an effort to further understand how our models perform on the whole galaxy sample, we investigate how the values of disc fraction obtained by our model depends on mass and color; as a reminder, here `disc fraction' refers to the mass fraction of stars that are in the disc.
We explore how disc fraction changes with mass. In Fig.~\ref{fstar_vs_mass}, we show 2D histograms of the values of disc fraction versus mass, with marginal distributions shown on the sides.  The three panels show results for the three different methods of estimating the disc fraction.
The first thing  to note here is the absence of obvious bimodality in any of the disc fraction distributions at fixed mass. For this reason, the separation between bulge-dominated and disc-dominated galaxies is non-trivial. However, for the 1D and 2D models we consider galaxies that have more than half of their mass in the disc to be disc-dominated, 
hence we employ an (somewhat arbitrary, but popular in the literature) cut of 0.5 to identify the disc-dominated population. For consistency, we will be comparing our results with observational data that define disc-dominated populations with the same cut, but through the use of light profile decomposition.  
 The main trend shown on all three panels is that the  disc-dominated population is concentrated between $\log_{10}(M_*/M_\odot)=10$ and $ \log_{10}(M_*/M_\odot)=10.5 $; 
 and the bulge dominated population spans the whole range of mass.
We also investigated color-dependence of the disc fractions, since from observations we know that morphology and color are strongly correlated \citep{sdss_color-morph_2001}.  
Fig.~\ref{g-r-color} shows disc fraction calculated by the Monte Carlo method against the $g-r$ rest-frame color. 
Evidently, at $g-r \sim 0.7 $ and high $f^\text{disc}_\text{2D}$ 
are the red disc-dominated population which is more abundant than expected. This excess amount may be attributed to IllustrisTNG100 producing more red galaxies at $\log_{10}(M_*/M_\odot)=10.5 - 11.5$  range compared to the real Universes, as evidenced by the SDSS data \citep{tng-bimodal}. 
The 1D histogram in $g-r$ also reveals that red discs are more common than red bulge galaxies at $g-r\sim 0.7$. 
Fortunately, this unphysical aspect of the color-morphology relation does not substantially affect our study, which is focused primarily on morphology alone.  We therefore focus primarily on validation of the morphologies in the remaining subsections. 

\subsection{Thin disc, thick disc and counter-rotating disc}\label{thin-thick-section}

\begin{figure}
 \includegraphics[width=\columnwidth]{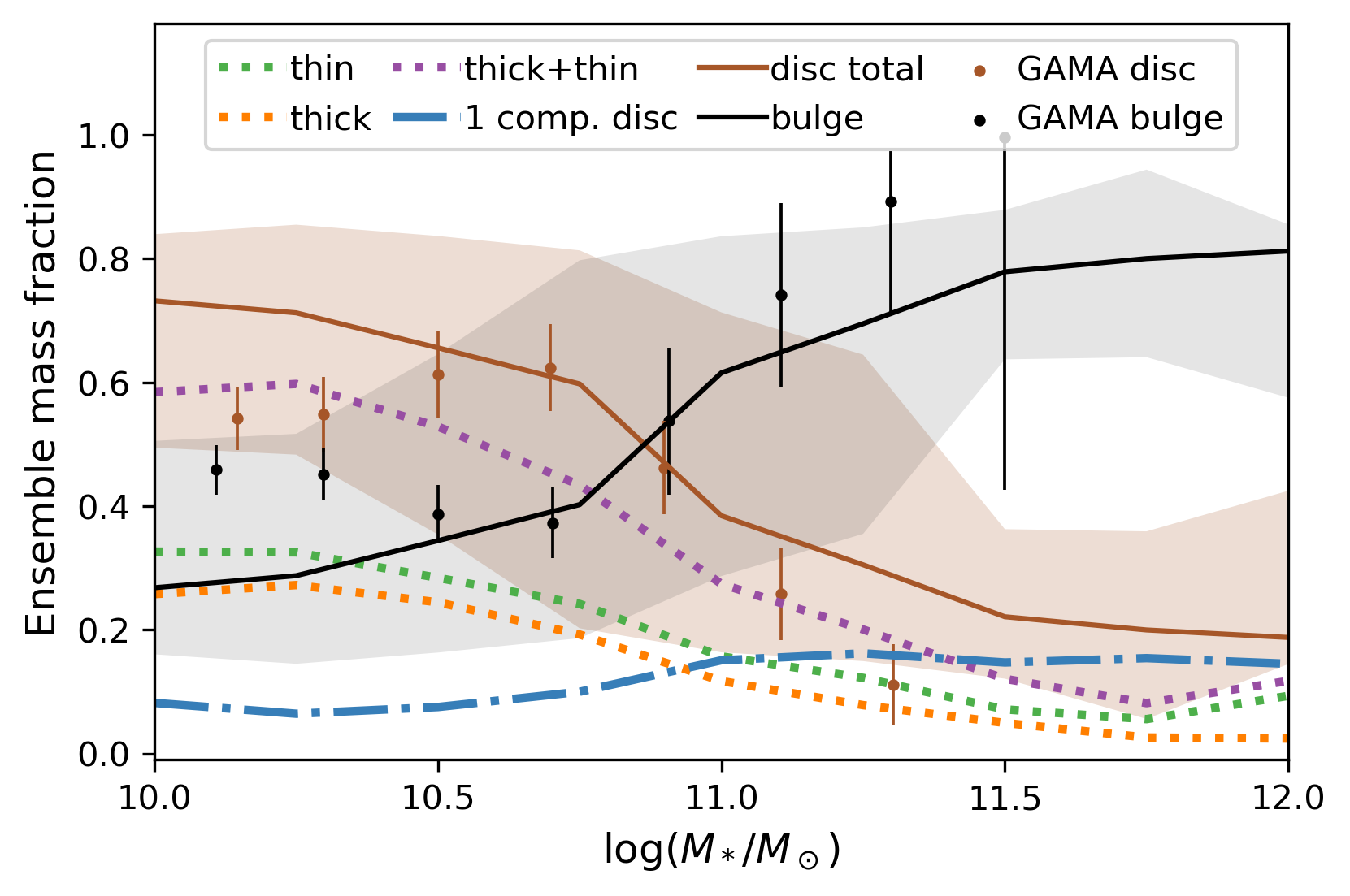}
 \caption{\label{total_mass_budget}  
 Contribution of galaxy components to the total stellar mass budget at a given stellar mass. The solid lines represent the fractional contribution for a given component, while solid colors represent the 16th and the 84th percentiles of the population.  
 The bulge and total disc fraction is compared with observational data from GAMA, with errorbars representing uncertainties on the mean. 
 The  dot-dashed blue line represents the contribution from those galaxies chosen by AIC as having only one disc component;   
 similarly the purple dotted line represents the contribution from the subset of galaxies chosen by AIC as having two disc components. 
 Further, the two component model (purple dotted) is broken down into thin and thick disc contributions, in green and orange dotted lines, respectively. The counter-rotating disc component is not shown here, since its contribution to the total mass budget is negligible.  
 }
\end{figure}

In this subsection, we investigate the contribution from each structural component to the total mass budget of galaxies 
which is given by $ {\sum_i M_{\text{y},i}}/{\sum_i M_{*,i}}$  for $N$ galaxies indexed by $i$ inside a stellar mass bin.  Here $y=$ thin disc, thick disc, thin+thick disc (2 component disc), 1 component disc, total disc and bulge. These quantities are plotted in Fig.~\ref{total_mass_budget}, where we demonstrate the population scatter in the individual galaxy values of $M_{\text{y},i}/M_{*,i}$ for the total disc and bulge contribution with shaded regions from the 16th to the 84th percentiles.  
 As expected, at high mass, the bulge component dominates the mass budget, while the disc structure dominates at low mass. The curves cross at $\log_{10}(M_*/M_\odot) =10.8  $,  
when the two components contribute equally. For comparison we show a comparable quantity from GAMA survey observations \citep{gama} using a photometric bulge/disc decomposition. GAMA is a combined
spectroscopic and multi-wavelength imaging survey designed
to study both galaxies and large-scale structure. The simulations and observations exhibit similar trends.

At high mass the disc component does not approach the low value expected from the GAMA data.   
One contributing factor to this effect is that our dynamical model will always have some non-negligible fraction of particles pointing the direction of the total angular momentum of the galaxy.  
For example, we found that in a spherically symmetric dispersion-dominated system on average this effect is $\sim 3$ per cent. Additionally, as the ellipticity is increased and the dispersion is no longer the same along the three axes (to resemble a realistic bulge-dominated galaxy), this effect increases as well. However, these may not fully explain the large disc fraction at high mass. We suspect that this may indicate an over-production of massive disc structures within the simulation, possibly related to the the over-production of red disc-dominated galaxies.   
The one-disc component contributes fairly equally  $\sim 0.2$ at all mass ranges, except for a shallow dip at around $\log_{10}(M_*/M_\odot)=10.25$, where the thin and thick curves peak. Thin and thick disc structures follow the same trend starting high at high mass. Thus, our results qualitatively agrees with observational data with some minor differences at high mass.

\begin{figure}
 \includegraphics[width=\columnwidth]{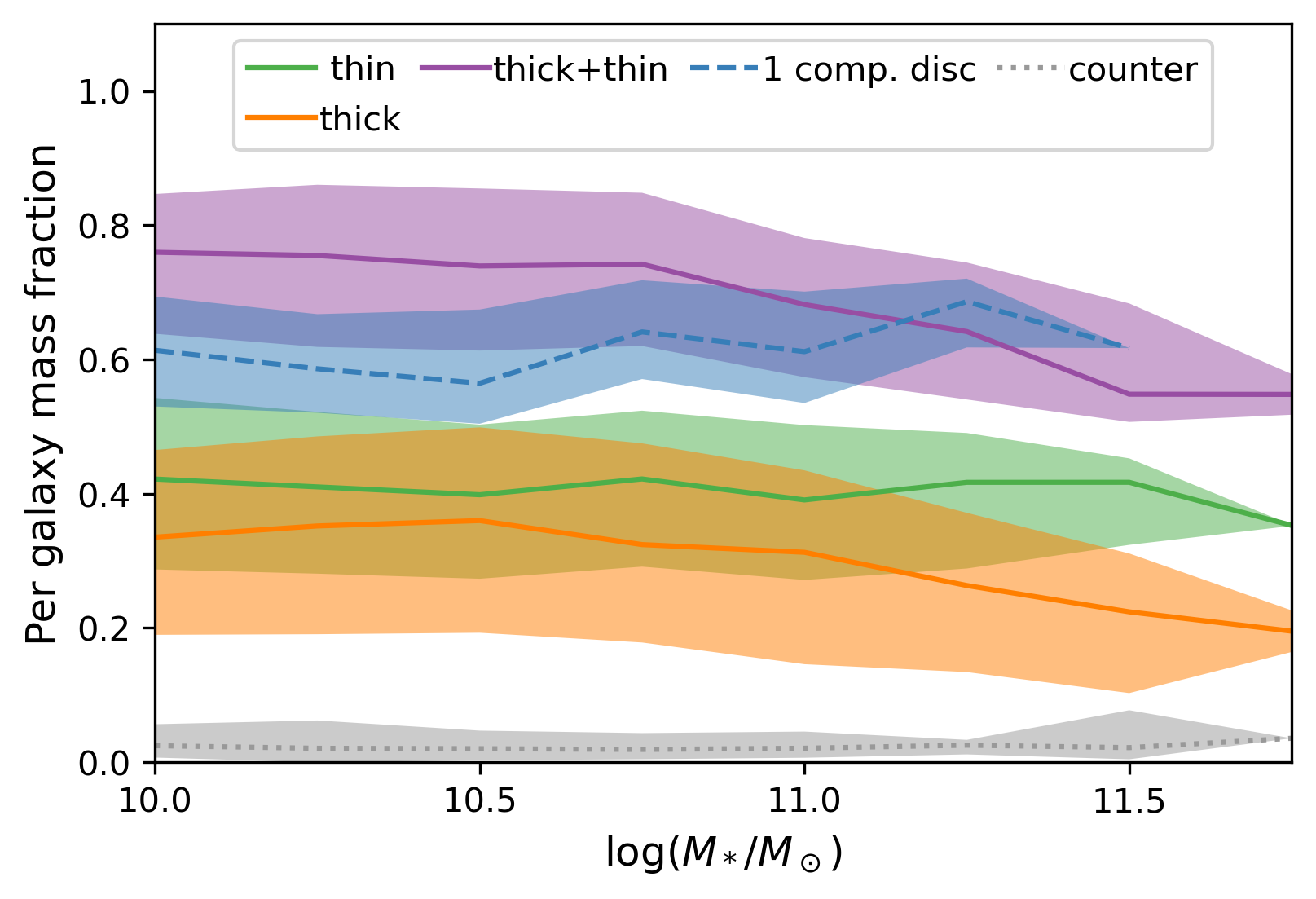}
 \caption{\label{thin_thick}  
 Mass fraction of disc components in disc-dominated galaxies, defined on a per-galaxy basis. The lines represent medians, while filled colors represent $1\sigma$ scatter in the population.   
 The dashed blue line represents the subset of galaxies identified as having a single disc component, while purple solid lines represent the total for galaxies with  two disc components. Further, the contributions of the thin and thick disc components are shown as green and orange lines, respectively, for the two-component galaxies. The dotted grey line indicates the counter-rotating disc fraction for galaxies that contain them (chosen by AIC).   
 }
\end{figure}

In Fig.~\ref{thin_thick} we present $M_{\text{y},i}/M_{*,i}$, the per galaxy mass fraction of various disc components in disc-dominated galaxies determined by the 1D model.  
The disc-related substructures identified through the 1D method all are found to contribute constant mass fraction over the range $ 10\le \log_{10}(M_*/M_\odot) \le 11.5  $. 
For galaxies with two disc components (thin and thick), the disc structure contributes roughly 70\% of the total mass budget.  At higher mass ($\log_{10}(M_*/M_\odot)$ > 11) it starts to exhibit a mild dependence on mass, decreasing to 0.6. 
The thick disc roughly makes up 30\% of the total mass budget and appears to be responsible for the decrease at higher mass. In contrast, the thin disc component stays at $\sim 0.4$ with no apparent dependence on mass.

For the 1D model, about $\sim $1000 galaxies (5\% of the whole sample) were chosen by AIC hosting a counter-rotating disc component. Interestingly, the counter-rotating disc fraction follows the same flat trend in stellar mass as did the other disc components, albeit contributing a very small fraction of less than 5\% to the total mass budget for the galaxies that host them. A very small number of galaxies ($\sim$70) include a  significant counter-rotating component with  20-30\% of their total stellar mass.
 
\subsection{Comparing surface profiles of bulges and discs against observational data}
 
In order to further explore the 2D model's efficacy, we fit the surface mass density 
profiles to parametric models, which enables us to quantify the features of these components for comparison with real data.   
We choose an arbitrary axis with reference to the simulation box, and project the 3D mass profiles for all galaxies onto the 2D plane.  We then construct profiles of the surface mass density in circular annuli about the galaxy centroid position.
For the whole sample we impose circular symmetry  on the surface mass density profiles and fit a 1D, two-component surface density profile: a S\'ersic profile for the bulge component and an exponential (i.e., S\'ersic $n=1$) profile for the disc component. 

\begin{figure*}
\begin{center}
$\begin{array}{c@{\hspace{0.5in}}c}
\includegraphics[width=7.2in]{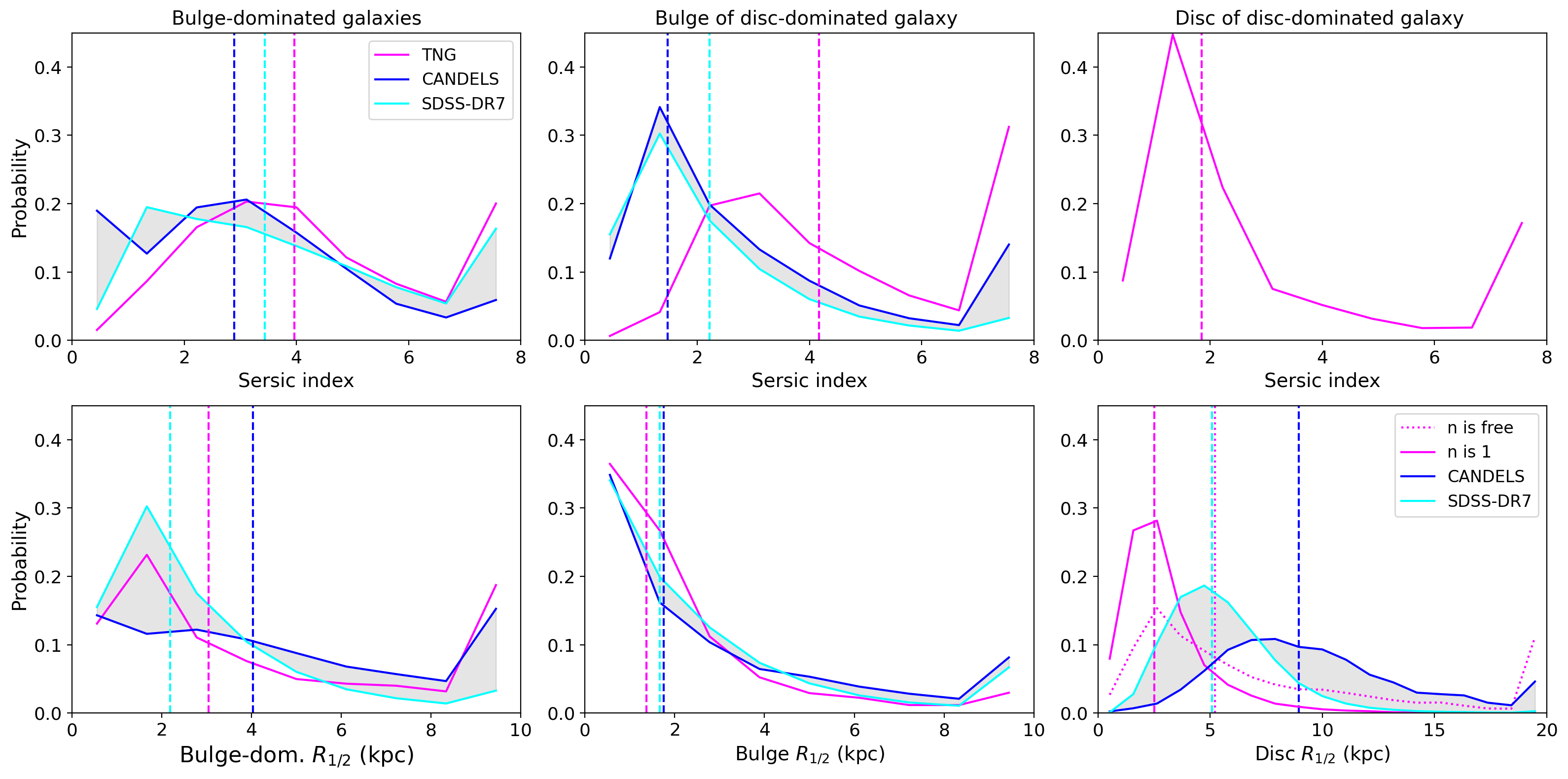} 
\end{array}$
\caption{ \label{S\'ersic 1} 
Distribution of S\'ersic  indices and half-mass radii of galaxy structural components identified using the 2D decomposition method  
in IllustrisTNG100-1 compared with CANDELS \protect\citep{CANDELS} and SDSS-DR7 
\protect\citep{meert2015}. The first column is for bulge-dominated galaxies 
only; the second column is for the bulge components of disc-dominated galaxies, and the third column is for the disc components of disc-dominated galaxies. The vertical dashed lines show the medians of each distribution. The grey shaded regions shows the range of values that lie between the two sets of observational results. 
Overall, for bulge-dominated galaxies the shapes of S\'ersic index distributions and the $R_{1/2}$ distributions have similar shapes and the medians for the S\'ersic indices are within 20\%. For the bulge component of disc-dominated galaxies, the bulge sizes agree both in shapes of the distributions and in medians to within 20\%, while the S\'ersic index distribution for TNG100 is shifted to the right compared to the two observational curves, with a median exceeding those in observations by a factor of $\sim$2.     
The S\'ersic  index distribution of disc components has a mode of 1, roughly consistent with expectations. However, the comparison of  sizes of disc components of all three samples differ.   
The disc sizes of TNG100 are smaller than both CANDELS and SDSS-DR7, which may indicate unrealistic signatures in the simulated disc population.
The S\'ersic  indices for the CANDELS and SDSS-DR7 catalog were constrained not to exceed $n=8$, whereas we did not place any constraints on this quantity when modeling the galaxies in TNG.  When plotting the distribution, all values greater than 8 were put into the last bin.  
 } 
\end{center}
\end{figure*}

 In Fig.~\ref{S\'ersic 1}, we compare the resulting fits with CANDELS data from HST  
 from the \cite{CANDELS} catalog and SDSS-DR7 data from the \cite{meert2015} catalog.   
 For the \cite{CANDELS} catalog, we select a volume-complete sample by taking  galaxies with $z<1$ and $ \log_{10}(M_*/M_\odot)>10$. 
 The \cite{meert2015} catalog has an atypical mass distribution due to selection criteria that are difficult to reproduce in the simulation, so we resample from the catalog, so that the mass distribution is similar to both IllustrisTNG100 and \cite{CANDELS} catalog. One should note that this is only an approximate way to generate a volume-complete sample because of the implicit assumption that the galaxies were removed randomly through a process that does not depend on any of their properties other than mass. 
 In this section we will focus on the 2D model, and use a BTR $< 0.5$ cut for the disc-dominated galaxies (BTR $ \geq 0.5$ cut for the bulge-dominated galaxies) both in the simulation and  in the observational datasets.  
 
The distribution of half-mass radii of bulge-dominated  
galaxies from TNG100 quantitatively compares well with SDSS-DR7 catalog, with modes coinciding at 1.9 kpc and with medians at 3 kpc and 2.2 kpc respectively.  The shapes of the distributions are similar as well, despite a longer tail for TNG100. The CANDELS catalog's curve follows a similar trend as TNG100's curve above 3 kpc, however  it differs below 3 kpc, where it is considerably flatter.   

The distribution of half-mass radii of  the bulge component for disc-dominated galaxies  
is also in good quantitative agreement with both observational datasets.  
All three distributions follow a similar shape. The distributions for SDSS-DR7 and CANDELS exhibit a longer tail, so their medians are located slightly to the right at $\sim 1.8$ kpc compared to the distribution for TNG100  at 1.5~kpc.

 The median S\'ersic  index of the bulge components for disc-dominated galaxies  
 for TNG100 using our model is at $n=4.1$, which is remarkably close to the  
 value of $n=4$ (De Vaucouleurs law), albeit with a very long tail. However, CANDELS and SDSS-DR7 have lower medians of $n=1.6$ and $n=2.1$ respectively, with the overall distributions shifted slightly to the left compared to TNG100.
Based on these results for the bulge components, we assert that our kinematically decomposed bulge components compare favorably with prior expectation for bulges (De Vaucouleurs law), despite the disagreement with photometric decompositions of real galaxies in S\'ersic index. Perhaps the disagreement could be explained by the challenges of measuring the bulge of a disc-dominated galaxy effectively in real observations, since the S\'ersic index comparison for bulge-dominated galaxies is in agreement both in distribution shape and median values within $\sim20\%$. 
To further test the effectiveness of our model, we fit a S\'ersic  profile to the disc component by setting $n$ free.  We could not find two-component S\'ersic  fits of real galaxies for which the disc component is not set to $n=1$ so as to compare with observations. However, the  mode is around 1 with a median of 1.9, which is not far off from the expected $n=1$. Additionally, the bulges and discs of disc-dominated galaxies have different S\'ersic index distributions which hints at the structural difference of the two components.  
 
Finally, we compare the sizes of the disc components obtained through our model against observational data. We calculated the half-mass radii for the IllustrisTNG100 sample using two methods: first, with the disc  S\'ersic  index set to $n=1$, and then with the disc  S\'ersic  index is set free. When fixing disc $n=1$, the size distribution is narrower, with median and mode at $\sim$2.5~kpc.  In contrast, when $n$ is free, the mode remains unchanged and the median approximately doubles to $\sim$5.2~kpc  
due to the long tail to the right.  
The median of disc half mass radii at 2.5~kpc in IllustrisTNG100 is smaller by a factor of 2 and 4 than the median in SDSS-DR7 and CANDELS data, respectively, due to an overall shift in the distribution. However, the median size of disc components in TNG100  increases when $n$ is free, and is similar to the SDSS-DR7 median at $\sim$ 5.2 kpc, though the mode is unchanged. The smaller sizes of discs in the simulation shown here is  broadly consistent with what was found in previous studies \citep{tngimage2019}.  
The similarities between the distributions of the sizes and S\'ersic indices of bulge-dominated galaxies (between the simulation and observations) suggest that the simulated galaxy population is fairly realistic and our model effectively identifies a bulge-dominated population.  
 
All in all, the half-mass radii of bulge components of disc-dominated galaxies compare well with those in the observations, despite the disagreement in their S\'ersic index distribution. Additionally, when the disc components are fit to a general S\'ersic function, the mode of the S\'ersic index distribution is at the expected value of 1.  
However, the characteristic disc sizes do differ; we suspect the difference arises from the simulation producing undersized discs.

\subsection{Exploring the probabilistic nature of the 2D model}\label{rsd_sec}
\begin{figure*} 
\begin{subfigure}{.3\textwidth}
\centering
\includegraphics[width=5.5cm]{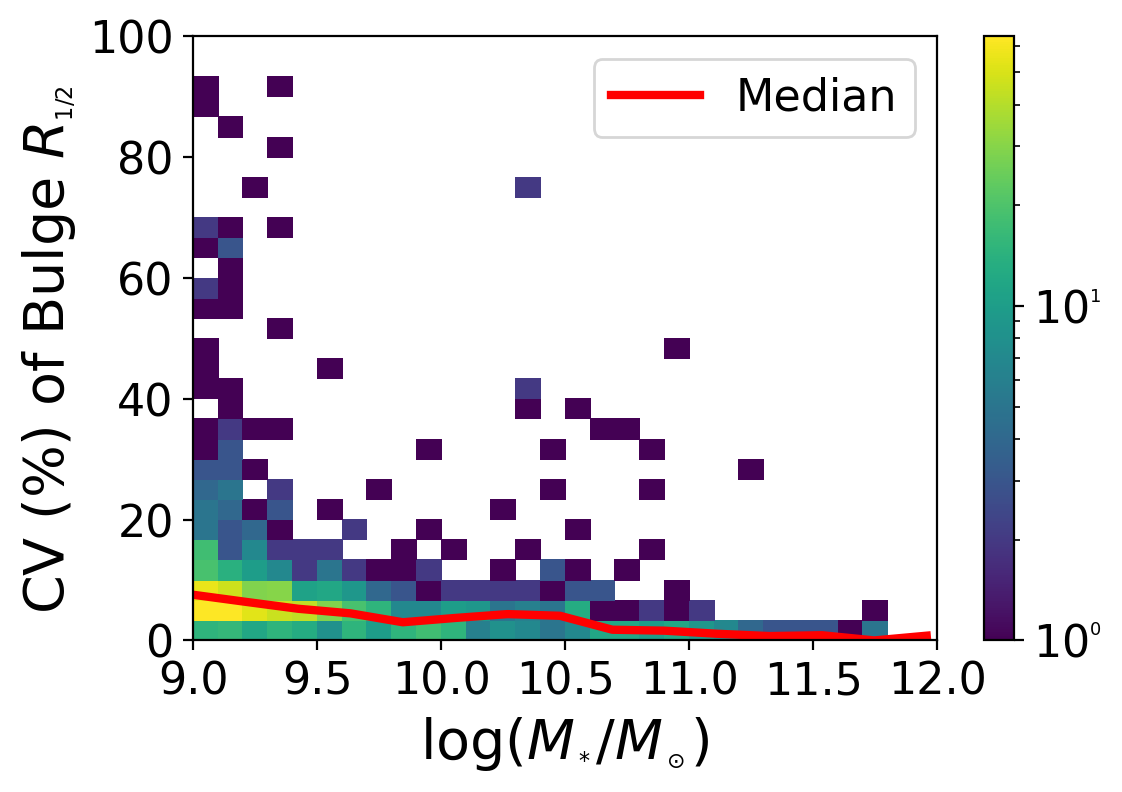}
\caption{ The CV of bulge $R_{1/2}$ for bulge-dominated galaxies in the subsample}
\end{subfigure}\hfill
\begin{subfigure}{.3\textwidth}
\centering
\includegraphics[width=5.5cm]{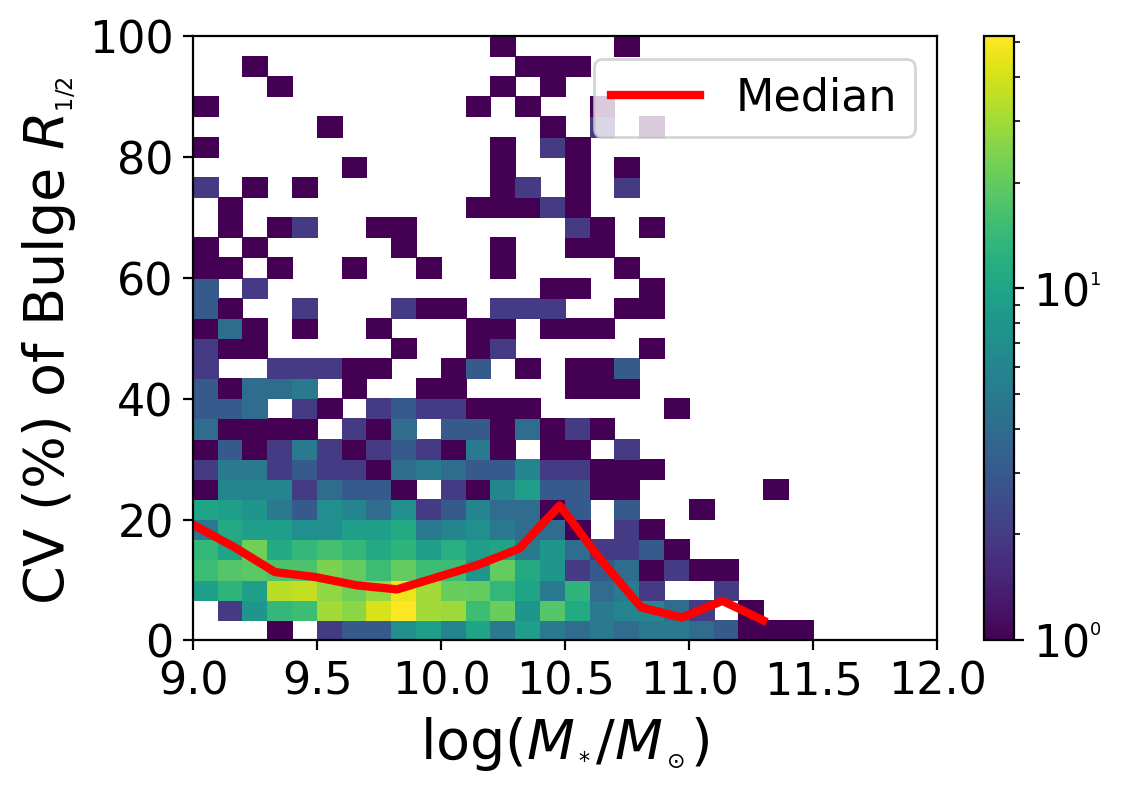}
\caption{ The CV of bulge $R_{1/2}$ for disc-dominated galaxies in the subsample}
\end{subfigure}\hfill
\begin{subfigure}{.3\textwidth}
\centering
\includegraphics[width=5.5cm]{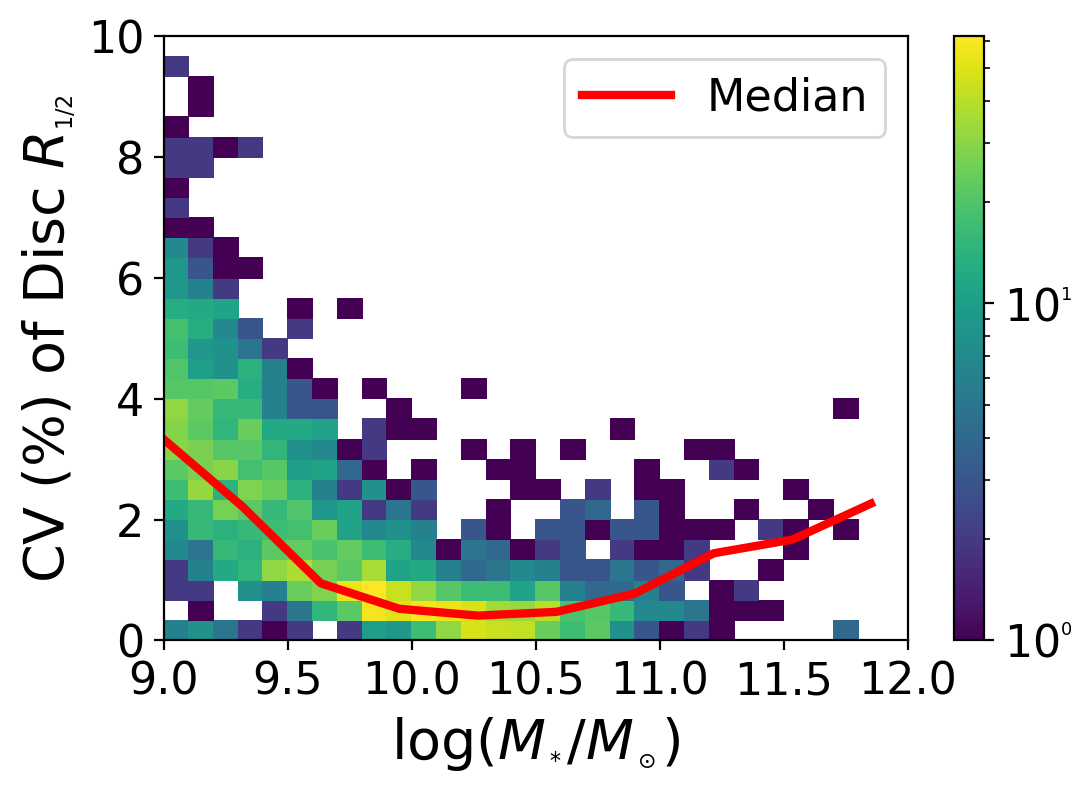}
\caption{ The CV of disc $R_{1/2}$ for all galaxies in the subsample}
\end{subfigure}\hfill
\begin{subfigure}{.3\textwidth}
\centering
\includegraphics[width=5.5cm]{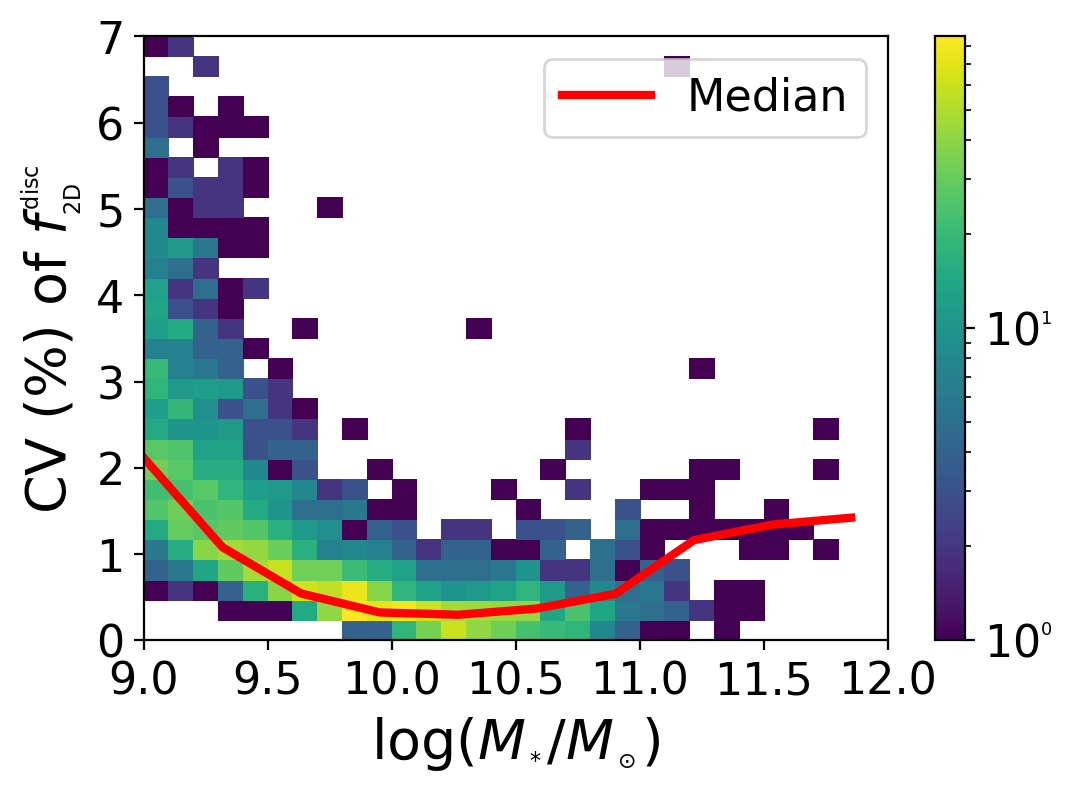}
\caption{ The CV of  $f_\text{2D}^\text{disc}$ for all galaxies in the subsample}
\end{subfigure}
\caption{\label{rsd}Distribution of relative uncertainty (CV), expressed in terms of per cent, of half-mass radii and disc fraction for a randomly-selected subsample of 3000 galaxies ($\sim$16\% of the whole sample).  
The colorbar indicates the number of galaxies in each bin. In a) the CV of bulge $R_{1/2}$ for bulge-dominated galaxies are below 20\% for most galaxies with a long tail of high CV values for galaxies around $\log_{10}(M_*/M_\odot)$ = 9. In b) the CV of bulge $R_{1/2}$ for disc-dominated galaxies most galaxies have CV <20\%, but with the outlier with high CV distributed uniformly across stellar mass.
As seen in b) and c) the CV values are below 10\% for most galaxies, with decreasing trends with increasing mass.} 
\end{figure*}

 Finally, we explore the probabilistic nature of the 2D model defined in \S~\ref{2d-method}. We choose a random subsample of 3000 galaxies (about 16\% of the ensemble) and repeat the Monte Carlo process 200 times to derive the standard deviation of measured individual galaxy properties such as half-mass radii of bulge and disc components and the fraction of stellar particles in the disc. 
 In Fig.~\ref{rsd} we present the relative uncertainty or the coefficient  of variation (CV), 
 defined as the standard deviation divided by the mean value, of the aforementioned measured quantities as a function of the stellar mass. Generally, we should expect larger uncertainties for low mass galaxies since they contain fewer particles.
Also, we expect small relative uncertainties for the fraction of stellar particles in the disc since the few mislabeled star particles will not have a big effect on the total mass of the disc (because we already excluded galaxies with fewer than 1000 particles). However,  the size of the components is more sensitive to mislabeled star particles because the position of the particles determines the size of the components, so the relative uncertainty is expected to be higher.
 
 Subfigures~\ref{rsd} a) and b) show the bulge $R_{1/2}$ CV for bulge-dominated and disc-dominated galaxies separately. 
 Both exhibit outliers with high CV values. The relatively high CV values (above 10\%) 
 for bulge $R_{1/2}$ could be explained by the fact that a stellar particle in the disc (far away from the center) once in a while will get classified as a bulge particle and in the process over-estimating the size of the bulge. 
 In contrast, for c) which presents the disc $R_{1/2}$   and d) which presents the  $f_\text{2D}^\text{disc}$,  all the CV values are under 10\%. Both exhibit decreasing trends with increasing mass, flattening out at around $\log_{10}(M_*/M_\odot)$ = 10. Interestingly, the bulge $R_{1/2}$ for disc-dominated galaxies does not exhibit this trend, suggesting that the high CV values have more to do with the stochastic nature of the method, except above $\log_{10}(M_*/M_\odot)$ = 10.75 where there is a decreasing trend. 
 These trends of low relative uncertainties shown above $\log_{10}(M_*/M_\odot)$ = 10 further supports our mass cut imposed in \S~\ref{n_frac}.  Overall, the disc fractions and disc sizes are determined with small (few percent) uncertainties, while bulges sizes are typically have larger uncertainties.  
 
 
\section{Conclusions}\label{conc}
 
In this work, we have introduced two galaxy decomposition models based on dynamics of star particles: a) a 1D model that can identify up to four galaxy structural components - bulge, thin disc, thick disc, and counter-rotating disc; b) a 2D model that has one more dimension  
which we chose to make probabilistic. These models improve upon existing models used in the literature to study large populations of galaxies in terms of quality of decomposition.   
They also bridge the gap between detailed models of disc galaxies from high resolution zoom-in simulations and structural studies of large galaxy ensembles in cosmological hydrodynamic simulations. We developed and tested these models using the IllustrisTNG100-1 cosmological hydrodynamical simulation, which includes galaxies of diverse morphologies. 

Based on the distributions of dynamical parameters of individual star particles within their host galaxies, we focused on two parameters: 1) $\cos\alpha$ - the alignment of a star particle's angular momentum vector with the galaxy's total angular momentum vector; and the 2) $j_\text{r}$ - which quantifies the star particle's orbital shape. We introduced two methods of galaxy classification/decomposition and compared them with a commonly-used, pre-existing method in the literature \citep{Abadi_2003}. 
The first, simple 1D model based on $\cos\alpha$ can describe as many as four underlying galaxy substructures: thin disc, thick disc, bulge and counter-rotating disc.
 
The second method adds one more dimension ($j_\text{r}$) to the model; we chose to implement the model with a probabilistic assignment of each star particle's domain based on its dynamics. 
Following the definition of our dynamical decomposition methods, we applied them to galaxy samples in IllustrisTNG100-1. When we explored the probabilistic aspect of the model, we found decreasing trends in uncertainties of measured quantities with increasing mass. The measured disc fractions and disc sizes had for the most part uncertainties less than 10\%, whereas uncertainties in the bulge sizes were higher.

Given our dynamical decompositions for the simulated galaxies, we first investigated the dependence of the fraction of disc-dominated galaxies on stellar mass. Our results are in agreement with observed values \citep{conselice, bluck-sdss}  
to within $\sim 10 \%$ down to a stellar mass of $ \log_{10}(M_*/M_\odot)=10$, below which we believe the results differ due to resolution limitations of the simulation.  
When exploring the simulated rest-frame colors of the galaxies, we found two separate populations as expected (red and blue galaxies); however, there was an excess of red disc-dominated galaxies compared to expectations \citep{tng-bimodal}. 

Next, we investigated how various components of the disc structure and the bulge  
contribute to the total stellar mass budget  
at a given stellar mass. For disc-dominated galaxies, we found that contributions from the disc  structures do not depend on stellar mass. However, for the whole galaxy ensemble, the disc structure dominates the mass budget at low stellar mass, while the bulge dominates at higher mass, as expected.

Finally, we compared our results to photometrically decomposed galaxies in SDSS \citep{meert2015} and HST \citep{CANDELS}. 
The 1D surface density results reflect agreement with observations in some cases, but there are some features that may indicate limitations in the simulations and/or the decomposition methods. The half mass radii and S\'ersic  indices for bulges and bulge-dominated  
galaxies are in good agreement. The distribution of S\'ersic  indices for the disc components peaks at the expected value of 1. Nevertheless, the physical sizes of disc components are smaller than the sizes of observed disc components.  
One caveat of our dynamical models is that halo stars were not separated from the bulge component because halo stars are a small fraction of the mass compared to more prominent structures such as discs and bulges \citep{stellar_halo}.  

In future studies, we look to use this model to study the statistical properties of individual galaxy components. One such property is the intrinsic alignment, a key source of systematic bias in cosmological weak lensing measurements \citep{2016MNRAS.456..207K}. 
Previous studies of intrinsic alignments in simulations have used simulated galaxy samples without any morphological decomposition \citep[e.g.,][]{horizon-ia, mbii-ia,illustris-ia}, 
or have separated galaxy populations into disc and bulge populations using a single-parameter cutoff scheme \citep[e.g.,][]{Tenneti_2016}. Our dynamical model for decomposing large simulated galaxy ensembles robustly into substructures opens up the possibility of improved analysis of this and other physical effects involving individual galaxy components in cosmological hydrodynamic simulations. 

\section*{Acknowledgements}

We thank Simon Samuroff, Aklant Bhowmick, Sukhdeep Singh, Scott Dodelson for useful discussion that informed the direction of this work.  
This work was supported in part by the National Science Foundation, NSF AST-1716131.
\section*{Data Availability }
 The data used in this paper is publicly available.
The IllustrisTNG data can be obtained through the website at
\url{https://www.tng-project.org/data/}. The catalog data with morphological decompositions of galaxies in a single redshift snapshot for IllustrisTNG-100, and the example code to produce it, are available at \url{https://github.com/McWilliamsCenter/gal_decomp_paper}.

\bibliographystyle{mnras}
\bibliography{example}  

\begin{thebibliography}{}
\makeatletter
\relax
\def\mn@urlcharsother{\let\do\@makeother \do\$\do\&\do\#\do\^\do\_\do\%\do\~}
\def\mn@doi{\begingroup\mn@urlcharsother \@ifnextchar [ {\mn@doi@}
  {\mn@doi@[]}}
\def\mn@doi@[#1]#2{\def\@tempa{#1}\ifx\@tempa\@empty \href
  {http://dx.doi.org/#2} {doi:#2}\else \href {http://dx.doi.org/#2} {#1}\fi
  \endgroup}
\def\mn@eprint#1#2{\mn@eprint@#1:#2::\@nil}
\def\mn@eprint@arXiv#1{\href {http://arxiv.org/abs/#1} {{\tt arXiv:#1}}}
\def\mn@eprint@dblp#1{\href {http://dblp.uni-trier.de/rec/bibtex/#1.xml}
  {dblp:#1}}
\def\mn@eprint@#1:#2:#3:#4\@nil{\def\@tempa {#1}\def\@tempb {#2}\def\@tempc
  {#3}\ifx \@tempc \@empty \let \@tempc \@tempb \let \@tempb \@tempa \fi \ifx
  \@tempb \@empty \def\@tempb {arXiv}\fi \@ifundefined
  {mn@eprint@\@tempb}{\@tempb:\@tempc}{\expandafter \expandafter \csname
  mn@eprint@\@tempb\endcsname \expandafter{\@tempc}}}

\bibitem[\protect\citeauthoryear{Abadi, Navarro, Steinmetz  \& Eke}{Abadi
  et~al.}{2003}]{Abadi_2003}
Abadi M.~G.,  Navarro J.~F.,  Steinmetz M.,   Eke V.~R.,  2003, \mn@doi [The
  Astrophysical Journal] {10.1086/378316}, 597, 21

\bibitem[\protect\citeauthoryear{{Akaike}}{{Akaike}}{1974}]{Akaike}
{Akaike} H.,  1974, IEEE Transactions on Automatic Control, \href
  {https://ui.adsabs.harvard.edu/abs/1974ITAC...19..716A} {19, 716}

\bibitem[\protect\citeauthoryear{Bamford et~al.,}{Bamford
  et~al.}{2009}]{zoo_env}
Bamford S.~P.,  et~al., 2009, \mn@doi [Monthly Notices of the Royal
  Astronomical Society] {10.1111/j.1365-2966.2008.14252.x}, 393, 1324

\bibitem[\protect\citeauthoryear{{Bignone}, {Pedrosa}, {Trayford}, {Tissera}
  \& {Pellizza}}{{Bignone} et~al.}{2020}]{eagle2020}
{Bignone} L.~A.,  {Pedrosa} S.~E.,  {Trayford} J.~W.,  {Tissera} P.~B.,
  {Pellizza} L.~J.,  2020, \mn@doi [\mnras] {10.1093/mnras/stz3014}, \href
  {https://ui.adsabs.harvard.edu/abs/2020MNRAS.491.3624B} {491, 3624}

\bibitem[\protect\citeauthoryear{{Bluck} et~al.,}{{Bluck}
  et~al.}{2019}]{bluck-sdss}
{Bluck} A. F.~L.,  et~al., 2019, \mn@doi [\mnras] {10.1093/mnras/stz363}, \href
  {https://ui.adsabs.harvard.edu/abs/2019MNRAS.485..666B} {485, 666}

\bibitem[\protect\citeauthoryear{Brook et~al.,}{Brook et~al.}{2012}]{brook2012}
Brook C.~B.,  et~al., 2012, \mn@doi [Monthly Notices of the Royal Astronomical
  Society] {10.1111/j.1365-2966.2012.21738.x}, 426, 690

\bibitem[\protect\citeauthoryear{{Chisari} et~al.,}{{Chisari}
  et~al.}{2015}]{horizon-ia}
{Chisari} N.,  et~al., 2015, \mn@doi [\mnras] {10.1093/mnras/stv2154}, \href
  {https://ui.adsabs.harvard.edu/abs/2015MNRAS.454.2736C} {454, 2736}

\bibitem[\protect\citeauthoryear{{Conselice}}{{Conselice}}{2006}]{conselice}
{Conselice} C.~J.,  2006, \mn@doi [\mnras] {10.1111/j.1365-2966.2006.11114.x},
  \href {https://ui.adsabs.harvard.edu/abs/2006MNRAS.373.1389C} {373, 1389}

\bibitem[\protect\citeauthoryear{Croft, Di~Matteo, Springel  \&
  Hernquist}{Croft et~al.}{2009}]{croft2009}
Croft R. A.~C.,  Di~Matteo T.,  Springel V.,   Hernquist L.,  2009, \mn@doi
  [Monthly Notices of the Royal Astronomical Society]
  {10.1111/j.1365-2966.2009.15446.x}, 400, 43

\bibitem[\protect\citeauthoryear{{Dalcanton} \& {Bernstein}}{{Dalcanton} \&
  {Bernstein}}{2002}]{thickdisk}
{Dalcanton} J.~J.,  {Bernstein} R.~A.,  2002, \mn@doi [\aj] {10.1086/342286},
  \href {https://ui.adsabs.harvard.edu/abs/2002AJ....124.1328D} {124, 1328}

\bibitem[\protect\citeauthoryear{{Davis}, {Efstathiou}, {Frenk}  \&
  {White}}{{Davis} et~al.}{1985}]{fof}
{Davis} M.,  {Efstathiou} G.,  {Frenk} C.~S.,   {White} S.~D.~M.,  1985,
  \mn@doi [\apj] {10.1086/163168}, \href
  {https://ui.adsabs.harvard.edu/abs/1985ApJ...292..371D} {292, 371}

\bibitem[\protect\citeauthoryear{{De Lucia}, {Muzzin}  \& {Weinmann}}{{De
  Lucia} et~al.}{2014}]{gal_problems}
{De Lucia} G.,  {Muzzin} A.,   {Weinmann} S.,  2014, \mn@doi [\nar]
  {10.1016/j.newar.2014.08.001}, \href
  {https://ui.adsabs.harvard.edu/abs/2014NewAR..62....1D} {62, 1}

\bibitem[\protect\citeauthoryear{Dimauro et~al.,}{Dimauro
  et~al.}{2018}]{CANDELS}
Dimauro P.,  et~al., 2018, \mn@doi [{Monthly Notices of the Royal Astronomical
  Society}] {10.1093/mnras/sty1379}, 478, 5410

\bibitem[\protect\citeauthoryear{Doménech-Moral, Martínez-Serrano,
  Domínguez-Tenreiro  \& Serna}{Doménech-Moral et~al.}{2012}]{domenech2012}
Doménech-Moral M.,  Martínez-Serrano F.~J.,  Domínguez-Tenreiro R.,   Serna
  A.,  2012, \mn@doi [Monthly Notices of the Royal Astronomical Society]
  {10.1111/j.1365-2966.2012.20534.x}, 421, 2510

\bibitem[\protect\citeauthoryear{Du, Ho, Zhao, Shi, Debattista, Hernquist  \&
  Nelson}{Du et~al.}{2019}]{Du_2019}
Du M.,  Ho L.~C.,  Zhao D.,  Shi J.,  Debattista V.~P.,  Hernquist L.,   Nelson
  D.,  2019, \mn@doi [The Astrophysical Journal] {10.3847/1538-4357/ab43cc},
  884, 129

\bibitem[\protect\citeauthoryear{Du, Ho, Debattista, Pillepich, Nelson, Zhao
  \& Hernquist}{Du et~al.}{2020}]{Du_2020}
Du M.,  Ho L.~C.,  Debattista V.~P.,  Pillepich A.,  Nelson D.,  Zhao D.,
  Hernquist L.,  2020, \mn@doi [The Astrophysical Journal]
  {10.3847/1538-4357/ab8fa8}, 895, 139

\bibitem[\protect\citeauthoryear{{Feng}, {Di Matteo}, {Croft}, {Tenneti},
  {Bird}, {Battaglia}  \& {Wilkins}}{{Feng} et~al.}{2015}]{bluetides2015}
{Feng} Y.,  {Di Matteo} T.,  {Croft} R.,  {Tenneti} A.,  {Bird} S.,
  {Battaglia} N.,   {Wilkins} S.,  2015, \mn@doi [\apjl]
  {10.1088/2041-8205/808/1/L17}, \href
  {https://ui.adsabs.harvard.edu/abs/2015ApJ...808L..17F} {808, L17}

\bibitem[\protect\citeauthoryear{{Guo} et~al.,}{{Guo}
  et~al.}{2009}]{van_den_cent_gal}
{Guo} Y.,  et~al., 2009, \mn@doi [\mnras] {10.1111/j.1365-2966.2009.15223.x},
  \href {https://ui.adsabs.harvard.edu/abs/2009MNRAS.398.1129G} {398, 1129}

\bibitem[\protect\citeauthoryear{Heymans et~al.,}{Heymans
  et~al.}{2013}]{heymans}
Heymans C.,  et~al., 2013, \mn@doi [Monthly Notices of the Royal Astronomical
  Society] {10.1093/mnras/stt601}, 432, 2433

\bibitem[\protect\citeauthoryear{{Hilbert}, {Xu}, {Schneider}, {Springel},
  {Vogelsberger}  \& {Hernquist}}{{Hilbert} et~al.}{2017}]{illustris-ia}
{Hilbert} S.,  {Xu} D.,  {Schneider} P.,  {Springel} V.,  {Vogelsberger} M.,
  {Hernquist} L.,  2017, \mn@doi [\mnras] {10.1093/mnras/stx482}, \href
  {https://ui.adsabs.harvard.edu/abs/2017MNRAS.468..790H} {468, 790}

\bibitem[\protect\citeauthoryear{Hirata \& Seljak}{Hirata \&
  Seljak}{2004}]{hirata2004}
Hirata C.~M.,  Seljak U. c.~v.,  2004, \mn@doi [Phys. Rev. D]
  {10.1103/PhysRevD.70.063526}, 70, 063526

\bibitem[\protect\citeauthoryear{Hirata, Mandelbaum, Ishak, Seljak, Nichol,
  Pimbblet, Ross  \& Wake}{Hirata et~al.}{2007}]{hirata2007}
Hirata C.~M.,  Mandelbaum R.,  Ishak M.,  Seljak U.,  Nichol R.,  Pimbblet
  K.~A.,  Ross N.~P.,   Wake D.,  2007, \mn@doi [Monthly Notices of the Royal
  Astronomical Society] {10.1111/j.1365-2966.2007.12312.x}, 381, 1197

\bibitem[\protect\citeauthoryear{{Huang}, {Di Matteo}, {Bhowmick}, {Feng}  \&
  {Ma}}{{Huang} et~al.}{2018}]{bluetides}
{Huang} K.-W.,  {Di Matteo} T.,  {Bhowmick} A.~K.,  {Feng} Y.,   {Ma} C.-P.,
  2018, \mn@doi [\mnras] {10.1093/mnras/sty1329}, \href
  {https://ui.adsabs.harvard.edu/abs/2018MNRAS.478.5063H} {478, 5063}

\bibitem[\protect\citeauthoryear{Joachimi et~al.}{Joachimi
  et~al.}{2015}]{joachimi_2015}
Joachimi B.,  et~al., 2015, \mn@doi [Space Sci. Rev.]
  {10.1007/s11214-015-0177-4}, 193, 1

\bibitem[\protect\citeauthoryear{{Kiessling} et~al.,}{{Kiessling}
  et~al.}{2015}]{2015SSRv..193...67K}
{Kiessling} A.,  et~al., 2015, \mn@doi [\ssr] {10.1007/s11214-015-0203-6},
  \href {https://ui.adsabs.harvard.edu/abs/2015SSRv..193...67K} {193, 67}

\bibitem[\protect\citeauthoryear{{Kormendy} \& {Kennicutt}}{{Kormendy} \&
  {Kennicutt}}{2004}]{pseudobulge}
{Kormendy} J.,  {Kennicutt} Robert~C. J.,  2004, \mn@doi [\araa]
  {10.1146/annurev.astro.42.053102.134024}, \href
  {https://ui.adsabs.harvard.edu/abs/2004ARA&A..42..603K} {42, 603}

\bibitem[\protect\citeauthoryear{{Krause}, {Eifler}  \& {Blazek}}{{Krause}
  et~al.}{2016}]{2016MNRAS.456..207K}
{Krause} E.,  {Eifler} T.,   {Blazek} J.,  2016, \mn@doi [\mnras]
  {10.1093/mnras/stv2615}, \href
  {https://ui.adsabs.harvard.edu/abs/2016MNRAS.456..207K} {456, 207}

\bibitem[\protect\citeauthoryear{{Lee} et~al.,}{{Lee}
  et~al.}{2013}]{determines_size}
{Lee} J.~H.,  et~al., 2013, \mn@doi [\apjl] {10.1088/2041-8205/762/1/L4}, \href
  {https://ui.adsabs.harvard.edu/abs/2013ApJ...762L...4L} {762, L4}

\bibitem[\protect\citeauthoryear{Liddle}{Liddle}{2007}]{liddle}
Liddle A.~R.,  2007, \mn@doi [Monthly Notices of the Royal Astronomical
  Society: Letters] {10.1111/j.1745-3933.2007.00306.x}, 377, L74

\bibitem[\protect\citeauthoryear{Marinacci, Pakmor  \& Springel}{Marinacci
  et~al.}{2014}]{marinacci}
Marinacci F.,  Pakmor R.,   Springel V.,  2014, \mn@doi [Mon. Not. Roy. Astron.
  Soc.] {10.1093/mnras/stt2003}, 437, 1750

\bibitem[\protect\citeauthoryear{Marinacci et~al.}{Marinacci
  et~al.}{2018}]{Marinacci2017illustristng}
Marinacci F.,  et~al., 2018, \mn@doi [Mon. Not. Roy. Astron. Soc.]
  {10.1093/mnras/sty2206}, 480, 5113

\bibitem[\protect\citeauthoryear{{Meert}, {Vikram}  \& {Bernardi}}{{Meert}
  et~al.}{2015}]{meert2015}
{Meert} A.,  {Vikram} V.,   {Bernardi} M.,  2015, \mn@doi [\mnras]
  {10.1093/mnras/stu2333}, \href
  {https://ui.adsabs.harvard.edu/abs/2015MNRAS.446.3943M} {446, 3943}

\bibitem[\protect\citeauthoryear{{Merritt}, {van Dokkum}, {Abraham}  \&
  {Zhang}}{{Merritt} et~al.}{2016}]{stellar_halo}
{Merritt} A.,  {van Dokkum} P.,  {Abraham} R.,   {Zhang} J.,  2016, \mn@doi
  [\apj] {10.3847/0004-637X/830/2/62}, \href
  {https://ui.adsabs.harvard.edu/abs/2016ApJ...830...62M} {830, 62}

\bibitem[\protect\citeauthoryear{{Mo}, {van den Bosch}  \& {White}}{{Mo}
  et~al.}{2010}]{book}
{Mo} H.,  {van den Bosch} F.~C.,   {White} S.,  2010, {Galaxy Formation and
  Evolution}

\bibitem[\protect\citeauthoryear{Moffett et~al.,}{Moffett et~al.}{2016}]{gama}
Moffett A.~J.,  et~al., 2016, \mn@doi [Monthly Notices of the Royal
  Astronomical Society] {10.1093/mnras/stw1861}, 462, 4336

\bibitem[\protect\citeauthoryear{{Naab} \& {Ostriker}}{{Naab} \&
  {Ostriker}}{2017}]{theory_challenge}
{Naab} T.,  {Ostriker} J.~P.,  2017, \mn@doi [\araa]
  {10.1146/annurev-astro-081913-040019}, \href
  {https://ui.adsabs.harvard.edu/abs/2017ARA&A..55...59N} {55, 59}

\bibitem[\protect\citeauthoryear{{Naiman} et~al.,}{{Naiman}
  et~al.}{2018}]{Naiman2018illustristng}
{Naiman} J.~P.,  et~al., 2018, \mn@doi [\mnras] {10.1093/mnras/sty618}, \href
  {https://ui.adsabs.harvard.edu/abs/2018MNRAS.477.1206N} {477, 1206}

\bibitem[\protect\citeauthoryear{Nelson et~al.,}{Nelson
  et~al.}{2017}]{tng_2color}
Nelson D.,  et~al., 2017, \mn@doi [Monthly Notices of the Royal Astronomical
  Society] {10.1093/mnras/stx3040}, 475, 624

\bibitem[\protect\citeauthoryear{Nelson et~al.}{Nelson
  et~al.}{2018}]{tng-bimodal}
Nelson D.,  et~al., 2018, \mn@doi [Mon. Not. Roy. Astron. Soc.]
  {10.1093/mnras/stx3040}, 475, 624

\bibitem[\protect\citeauthoryear{{Nelson} et~al.,}{{Nelson}
  et~al.}{2019}]{tng-publicdata}
{Nelson} D.,  et~al., 2019, \mn@doi [Computational Astrophysics and Cosmology]
  {10.1186/s40668-019-0028-x}, \href
  {https://ui.adsabs.harvard.edu/abs/2019ComAC...6....2N} {6, 2}

\bibitem[\protect\citeauthoryear{Obreja, Macciò, Moster, Dutton, Buck, Stinson
   \& Wang}{Obreja et~al.}{2018}]{obreja2018}
Obreja A.,  Macciò A.~V.,  Moster B.,  Dutton A.~A.,  Buck T.,  Stinson G.~S.,
    Wang L.,  2018, \mn@doi [Monthly Notices of the Royal Astronomical Society]
  {10.1093/mnras/sty1022}, 477, 4915

\bibitem[\protect\citeauthoryear{{Oser}, {Naab}, {Ostriker}  \&
  {Johansson}}{{Oser} et~al.}{2012}]{size_velo_early}
{Oser} L.,  {Naab} T.,  {Ostriker} J.~P.,   {Johansson} P.~H.,  2012, \mn@doi
  [\apj] {10.1088/0004-637X/744/1/63}, \href
  {https://ui.adsabs.harvard.edu/abs/2012ApJ...744...63O} {744, 63}

\bibitem[\protect\citeauthoryear{{Park} \& {Choi}}{{Park} \&
  {Choi}}{2005}]{morph_color_color}
{Park} C.,  {Choi} Y.-Y.,  2005, \mn@doi [\apjl] {10.1086/499243}, \href
  {https://ui.adsabs.harvard.edu/abs/2005ApJ...635L..29P} {635, L29}

\bibitem[\protect\citeauthoryear{{Pillepich} et~al.,}{{Pillepich}
  et~al.}{2018a}]{tng-methods}
{Pillepich} A.,  et~al., 2018a, \mn@doi [\mnras] {10.1093/mnras/stx2656}, \href
  {https://ui.adsabs.harvard.edu/abs/2018MNRAS.473.4077P} {473, 4077}

\bibitem[\protect\citeauthoryear{{Pillepich} et~al.,}{{Pillepich}
  et~al.}{2018b}]{pillepich2018illustristng}
{Pillepich} A.,  et~al., 2018b, \mn@doi [\mnras] {10.1093/mnras/stx3112}, \href
  {https://ui.adsabs.harvard.edu/abs/2018MNRAS.475..648P} {475, 648}

\bibitem[\protect\citeauthoryear{Rodriguez-Gomez et~al.,}{Rodriguez-Gomez
  et~al.}{2018}]{tngimage2019}
Rodriguez-Gomez V.,  et~al., 2018, \mn@doi [Monthly Notices of the Royal
  Astronomical Society] {10.1093/mnras/sty3345}, 483, 4140

\bibitem[\protect\citeauthoryear{Sales, Navarro, Schaye, Vecchia, Springel  \&
  Booth}{Sales et~al.}{2010}]{sales_2010}
Sales L.~V.,  Navarro J.~F.,  Schaye J.,  Vecchia C.~D.,  Springel V.,   Booth
  C.~M.,  2010, \mn@doi [Monthly Notices of the Royal Astronomical Society]
  {10.1111/j.1365-2966.2010.17391.x}, 409, 1541

\bibitem[\protect\citeauthoryear{Scannapieco, White, Springel  \&
  Tissera}{Scannapieco et~al.}{2009}]{Scannapieco_2009}
Scannapieco C.,  White S. D.~M.,  Springel V.,   Tissera P.~B.,  2009, \mn@doi
  [Monthly Notices of the Royal Astronomical Society]
  {10.1111/j.1365-2966.2009.14764.x}, 396, 696–708

\bibitem[\protect\citeauthoryear{{Scherrer}, {Berlind}, {Mao}  \&
  {McBride}}{{Scherrer} et~al.}{2010}]{gauss-copula1}
{Scherrer} R.~J.,  {Berlind} A.~A.,  {Mao} Q.,   {McBride} C.~K.,  2010,
  \mn@doi [\apjl] {10.1088/2041-8205/708/1/L9}, \href
  {https://ui.adsabs.harvard.edu/abs/2010ApJ...708L...9S} {708, L9}

\bibitem[\protect\citeauthoryear{Sklar}{Sklar}{1959}]{Skla59}
Sklar A.,  1959, Publications de l'Institut de Statistique de l'Universit\'e de
  Paris, 8, 229

\bibitem[\protect\citeauthoryear{Somerville \& Davé}{Somerville \&
  Davé}{2015}]{Somerville2014}
Somerville R.~S.,  Davé R.,  2015, \mn@doi [Ann. Rev. Astron. Astrophys.]
  {10.1146/annurev-astro-082812-140951}, 53, 51

\bibitem[\protect\citeauthoryear{{Springel}}{{Springel}}{2010}]{arepo}
{Springel} V.,  2010, \mn@doi [\mnras] {10.1111/j.1365-2966.2009.15715.x},
  \href {https://ui.adsabs.harvard.edu/abs/2010MNRAS.401..791S} {401, 791}

\bibitem[\protect\citeauthoryear{{Springel}, {White}, {Tormen}  \&
  {Kauffmann}}{{Springel} et~al.}{2001}]{subfind}
{Springel} V.,  {White} S. D.~M.,  {Tormen} G.,   {Kauffmann} G.,  2001,
  \mn@doi [\mnras] {10.1046/j.1365-8711.2001.04912.x}, \href
  {https://ui.adsabs.harvard.edu/abs/2001MNRAS.328..726S} {328, 726}

\bibitem[\protect\citeauthoryear{Springel et~al.}{Springel
  et~al.}{2018}]{Springel2017illustristng}
Springel V.,  et~al., 2018, \mn@doi [Mon. Not. Roy. Astron. Soc.]
  {10.1093/mnras/stx3304}, 475, 676

\bibitem[\protect\citeauthoryear{Steinmetz \& Navarro}{Steinmetz \&
  Navarro}{2002}]{hierarchical}
Steinmetz M.,  Navarro J.,  2002, New Astronomy, 7, 155

\bibitem[\protect\citeauthoryear{{Strateva} et~al.,}{{Strateva}
  et~al.}{2001}]{sdss_color-morph_2001}
{Strateva} I.,  et~al., 2001, \mn@doi [\aj] {10.1086/323301}, \href
  {https://ui.adsabs.harvard.edu/abs/2001AJ....122.1861S} {122, 1861}

\bibitem[\protect\citeauthoryear{{Tabor}, {Merrifield}, {Arag{\'o}n-Salamanca},
  {Fraser-McKelvie}, {Peterken}, {Smethurst}, {Drory}  \& {Lane}}{{Tabor}
  et~al.}{2019}]{tabor-manga}
{Tabor} M.,  {Merrifield} M.,  {Arag{\'o}n-Salamanca} A.,  {Fraser-McKelvie}
  A.,  {Peterken} T.,  {Smethurst} R.,  {Drory} N.,   {Lane} R.~R.,  2019,
  \mn@doi [\mnras] {10.1093/mnras/stz431}, \href
  {https://ui.adsabs.harvard.edu/abs/2019MNRAS.485.1546T} {485, 1546}

\bibitem[\protect\citeauthoryear{Tacchella et~al.,}{Tacchella
  et~al.}{2019}]{tacchella2019}
Tacchella S.,  et~al., 2019, \mn@doi [Monthly Notices of the Royal Astronomical
  Society] {10.1093/mnras/stz1657}, 487, 5416

\bibitem[\protect\citeauthoryear{{Takeuchi}}{{Takeuchi}}{2010}]{gauss-copula2}
{Takeuchi} T.~T.,  2010, \mn@doi [\mnras] {10.1111/j.1365-2966.2010.16778.x},
  \href {https://ui.adsabs.harvard.edu/abs/2010MNRAS.406.1830T} {406, 1830}

\bibitem[\protect\citeauthoryear{{Taylor}, {Franx}, {Brinchmann}, {van der Wel}
   \& {van Dokkum}}{{Taylor} et~al.}{2010}]{mass_local}
{Taylor} E.~N.,  {Franx} M.,  {Brinchmann} J.,  {van der Wel} A.,   {van
  Dokkum} P.~G.,  2010, \mn@doi [\apj] {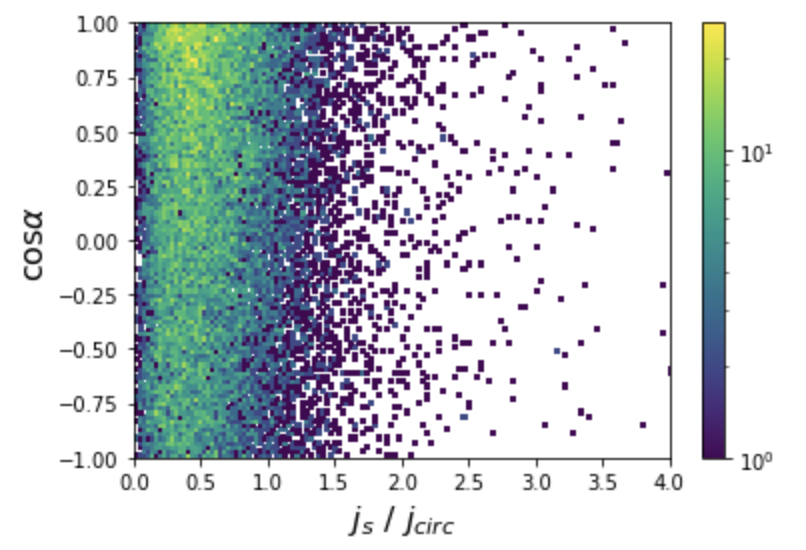}, \href
  {https://ui.adsabs.harvard.edu/abs/2010ApJ...722....1T} {722, 1}

\bibitem[\protect\citeauthoryear{{Tenneti}, {Singh}, {Mandelbaum}, {di Matteo},
  {Feng}  \& {Khandai}}{{Tenneti} et~al.}{2015}]{mbii-ia}
{Tenneti} A.,  {Singh} S.,  {Mandelbaum} R.,  {di Matteo} T.,  {Feng} Y.,
  {Khandai} N.,  2015, \mn@doi [\mnras] {10.1093/mnras/stv272}, \href
  {https://ui.adsabs.harvard.edu/abs/2015MNRAS.448.3522T} {448, 3522}

\bibitem[\protect\citeauthoryear{Tenneti, Mandelbaum  \& Di~Matteo}{Tenneti
  et~al.}{2016}]{Tenneti_2016}
Tenneti A.,  Mandelbaum R.,   Di~Matteo T.,  2016, \mn@doi [Monthly Notices of
  the Royal Astronomical Society] {10.1093/mnras/stw1823}, 462, 2668–2680

\bibitem[\protect\citeauthoryear{{Tortora}, {Napolitano}, {Cardone},
  {Capaccioli}, {Jetzer}  \& {Molinaro}}{{Tortora}
  et~al.}{2010}]{color_mass_stellar}
{Tortora} C.,  {Napolitano} N.~R.,  {Cardone} V.~F.,  {Capaccioli} M.,
  {Jetzer} P.,   {Molinaro} R.,  2010, \mn@doi [\mnras]
  {10.1111/j.1365-2966.2010.16938.x}, \href
  {https://ui.adsabs.harvard.edu/abs/2010MNRAS.407..144T} {407, 144}

\bibitem[\protect\citeauthoryear{Trujillo \& Bakos}{Trujillo \&
  Bakos}{2013}]{hst_halo}
Trujillo I.,  Bakos J.,  2013, \mn@doi [Mon. Not. Roy. Astron. Soc.]
  {10.1093/mnras/stt232}, 431, 1121

\bibitem[\protect\citeauthoryear{{Trujillo} et~al.,}{{Trujillo}
  et~al.}{2004}]{lum_size}
{Trujillo} I.,  et~al., 2004, \mn@doi [\apj] {10.1086/382060}, \href
  {https://ui.adsabs.harvard.edu/abs/2004ApJ...604..521T} {604, 521}

\bibitem[\protect\citeauthoryear{Vogelsberger et~al.,}{Vogelsberger
  et~al.}{2014}]{vogelsberger_2014}
Vogelsberger M.,  et~al., 2014, \mn@doi [Monthly Notices of the Royal
  Astronomical Society] {10.1093/mnras/stu1536}, 444, 1518

\bibitem[\protect\citeauthoryear{{Vogelsberger}, {Marinacci}, {Torrey}  \&
  {Puchwein}}{{Vogelsberger} et~al.}{2020}]{vogelsberger-review}
{Vogelsberger} M.,  {Marinacci} F.,  {Torrey} P.,   {Puchwein} E.,  2020,
  \mn@doi [Nature Reviews Physics] {10.1038/s42254-019-0127-2}, \href
  {https://ui.adsabs.harvard.edu/abs/2020NatRP...2...42V} {2, 42}

\bibitem[\protect\citeauthoryear{{Zhu}, {van de Ven}, {M{\'e}ndez-Abreu}  \&
  {Obreja}}{{Zhu} et~al.}{2018}]{zhu-califa}
{Zhu} L.,  {van de Ven} G.,  {M{\'e}ndez-Abreu} J.,   {Obreja} A.,  2018,
  \mn@doi [\mnras] {10.1093/mnras/sty1521}, \href
  {https://ui.adsabs.harvard.edu/abs/2018MNRAS.479..945Z} {479, 945}

\bibitem[\protect\citeauthoryear{de Vaucouleurs}{de~Vaucouleurs}{1948}]{de1948}
de Vaucouleurs G.,  1948, in Annales d'Astrophysique. p.~247

\bibitem[\protect\citeauthoryear{{de Vaucouleurs}}{{de
  Vaucouleurs}}{1958}]{de1958}
{de Vaucouleurs} G.,  1958, \mn@doi [\apj] {10.1086/146564}, \href
  {https://ui.adsabs.harvard.edu/abs/1958ApJ...128..465D} {128, 465}

\bibitem[\protect\citeauthoryear{{de Vaucouleurs}, {de Vaucouleurs}, {Corwin},
  {Buta}, {Paturel}  \& {Fouque}}{{de Vaucouleurs} et~al.}{1995}]{rc3}
{de Vaucouleurs} G.,  {de Vaucouleurs} A.,  {Corwin} H.~G.,  {Buta} R.~J.,
  {Paturel} G.,   {Fouque} P.,  1995, VizieR Online Data Catalog, \href
  {https://ui.adsabs.harvard.edu/abs/1995yCat.7155....0D} {p. VII/155}

\bibitem[\protect\citeauthoryear{{van der Kruit} \& {Freeman}}{{van der Kruit}
  \& {Freeman}}{2011}]{vanderKruit}
{van der Kruit} P.~C.,  {Freeman} K.~C.,  2011, \mn@doi [\araa]
  {10.1146/annurev-astro-083109-153241}, \href
  {https://ui.adsabs.harvard.edu/abs/2011ARA&A..49..301V} {49, 301}

\makeatother
\end{thebibliography}

\appendix

\bsp	
\label{lastpage}
\end{document}